\documentclass[11pt]{article}
\usepackage{times}
\usepackage[T1]{fontenc}
\usepackage{graphicx}
\usepackage{amssymb, amsmath}
\usepackage{rotating}
\renewcommand{\author}[1]{\large\rm #1\\ \bigskip}
\newcommand{\address}[1]{{\normalsize\it #1\\}\bigskip}
\renewcommand{\title}[1]{\bigskip\bigskip\Large\bf #1\bigskip\bigskip\\}

\usepackage{tikz}
\usetikzlibrary{shapes}
\def\root{\draw[fill] +(0,0) circle (5pt);}
\def\sww{\draw ++(0,0)}

\def\sdw{\draw[style=densely dotted] ++(0,0)}
\def\ew{;}
\def\r{-- ++(1,0)}
\def\l{-- ++(-1,0)}
\def\u{-- ++(0,1)}
\def\d{-- ++(0,-1)}

\usepackage{placeins}

\begin{document}
\begin{center}
\title{Three-dimensional terminally attached self-avoiding walks and bridges}
\author{Nathan Clisby,\footnote[1]{email:  {\tt
nclisby@unimelb.edu.au}  } Andrew R. Conway\footnote[2]{email:  {\tt
andrewscbridges@greatcactus.org}  }  and Anthony J.
Guttmann\footnote[3]{email: {\tt guttmann@unimelb.edu.au}}      }

\address{ ARC Centre of Excellence for Mathematics and Statistics of
Complex Systems,\\
School of Mathematics and Statistics,\\
The University of Melbourne, Victoria 3010, Australia}
\end{center}
\begin{abstract}
We study terminally attached self-avoiding walks and bridges on the simple cubic
lattice, both by series analysis and Monte Carlo methods. We provide
strong numerical evidence supporting a scaling relation between
self-avoiding walks, bridges, and terminally attached self-avoiding
walks, and posit that a corresponding amplitude
ratio is a universal quantity.

\end{abstract}
\section{Introduction}\label{Intro}
A self-avoiding walk (SAW) on a lattice is an open, connected path on
the lattice that does not revisit any previously visited vertex.  Walks
are considered distinct if they are not translates of one another. 
If we let the number of SAWs of $n$ steps be $c_n$, 
it is known that $\lim_{n\to\infty} n^{-1}  \log
c_n = \log \mu$ exists~\cite{HM54}, where $\mu$ is the 
\emph{growth constant} of self-avoiding walks on the lattice.
This work will consider SAWs on the
$3$-dimensional simple cubic lattice ${\mathbb Z}^3,$ with the vertices
having 
integer coordinates $\{x^{(i)},y^{(i)},  z^{(i)} \},$ for $i=0,1,
\cdots,n.$

The upper half-space ${\mathbb H},$ is characterised by 
$z \ge 0.$ An $n$-step \emph{bridge} is a self-avoiding walk in the
upper half-space that starts 
at the origin and is constrained so that $z^{(0)} < z^{(i)} \leq
z^{(n)}$ for any $0 < i \leq n$.
We denote the number of $n$-step bridges starting at the origin 
by $b_n$.  It is known that 
$\lim_{n\to\infty} n^{-1} \log b_n = \log \mu,$ where $\mu$ is unchanged
from the corresponding value for SAWs~\cite{HW62}.

A terminally attached self-avoiding walk,
or TAW, is a SAW with one end anchored in the
surface, but with the rest of the walk free in the upper half-space.
Terminally attached self-avoiding walks are also referred to as
half-space self-avoiding walks in the literature.
Clearly TAWs are a superset of bridges, and a subset of SAWs,
so have the same growth constant. The number of $n$-step TAWs
starting at 
the origin is denoted by $t_n.$

The last subset of SAWs we wish to consider are {\em arches}, which are
SAWs in the upper half-space with both the start- and end-point
constrained to lie in the $2$-dimensional surface $z=0$. That is to say,
$z^{(0)} = 0 = z^{(n)}.$ As the number of arches is bounded above by the
number of SAWs and below by the number of self-avoiding polygons (SAPs),
which are known to have the same growth constant as SAWs~\cite{H61}, it
follows that arches also have the same growth constant as SAWs. The
number of $n$-step arches starting at 
the origin is denoted by $a_n.$

The results on growth constants are the only results that have been
proved. Nevertheless, it is universally accepted that the asymptotic
behaviour of the above objects is given by:
\begin{align}
c_n &\sim A \cdot \mu^n \cdot n^{\gamma-1},
\label{eq:saw_asympt} \\
b_n &\sim B \cdot \mu^n \cdot n^{\gamma_b-1},
\label{eq:bridge_asympt} \\
t_n &\sim   H \cdot\mu^n \cdot n^{\gamma_1-1},
\label{eq:t_asympt} \\
a_n &\sim  C \cdot\mu^n \cdot n^{\gamma_{11}-1},
\label{eq:arch_asympt}
\end{align}
for SAWs, bridges, TAWs and arches respectively. The
existence of the various critical exponents has not been proved, but is
universally accepted, and will be assumed hereinafter.

All the above definitions also hold for the two-dimensional square
lattice, where the upper half-space referred to becomes the upper
half-plane, and the originating surface is the line $y=0.$

Another critical exponent that needs to be defined is that
which characterises the length of a SAW. All standard measures of the
square of the length,
such as the mean-squared end-to-end distance, the radius of gyration,
the squared caliper span etc.,  behave as ${\rm const.} \times n^{2\nu},$
where in two dimensions it is accepted that $\nu = 3/4.$ This exponent can also be understood as
the reciprocal of the fractal dimension, $d_f.$

In two dimensions, the exponents are known exactly (assuming existence).
In that case it is believed that $\gamma = 43/32,$ $\gamma_1 =
61/64$, and $\gamma_{11} = -3/16.$ As far as we are aware, there are
no published estimates for $\gamma_b,$ but as long ago as last millennium
one of us (AJG) estimated the value of this exponent by series analysis
to be $9/16.$ Subsequently, much longer series have been produced by
Iwan Jensen, which enabled this estimate to be affirmed with much
greater confidence. Somewhat later, Alberts and Madras (private
communication) obtained the estimate $\gamma_b=9/16$ for two-dimensional
bridges through SLE arguments, subject to certain unproven assumptions,
but this work was never published.

In~\cite{DGK11} both SAWs spanning a strip and bridges were discussed,
and comparisons made with conjectured results from ${\rm SLE}_{8/3}.$ The arguments
given there allow one to predict $\gamma_b,$ and they are reproduced, in summary, here.
In~\cite{LSW04} it was explained why the measure of bridges starting at the origin and ending
at  $x + iy$ should be $y^{-5/4} \, f(x/y)$  for an (explicit) function
$f$ that decays exponentially. Summing over all values of $x,$ it follows that the
measure of bridges starting at the origin and ending at height $h$ should be asymptotic 
to a constant times $h^{-1/4}.$ Then by integrating over height,
the measure of bridges  of height less than or equal to $h$ should grow as $h^{3/4}.$

Recall that for two-dimensional SAWs and bridges the exponent
$\nu=1/d_f=3/4.$ If we weight each bridge of length $n$ by
$\mu^{-n}$, so
that the normalised counts are given by $b_n \mu^{-n}$,
it follows that normalised bridges of height $h$ typically have
$h^{d_f}=h^{4/3}$ steps. Hence the measure of bridges of $n$ steps
should be the same as the measure of
bridges of height at most $n^{3/4}$,
which gives $(n^{3/4})^{3/4} = n^{9/16}$ for the measure. Equivalently, the number
of bridges of exactly $n$ steps grows like $b_n \sim const \times \mu^n
\times n^{9/16-1} = {\rm const.} \times \mu^n \times n^{-7/16},$
so that $\gamma_b =9/16.$

These exponents are not all independent. There is a scaling relation,
due to Barber~\cite{B73},
\begin{align}
2\gamma_1 - \gamma_{11} &= \gamma + \nu,
\label{eq:scalinga}
\end{align}
which holds independent of
dimension. Very recently, Duplantier and Guttmann~\cite{DG15} have
proposed the existence of another scaling relation giving the bridge
exponent,  
\begin{align}
\gamma_b & =\gamma_{11}+\nu.
\label{eq:scalingb}
\end{align}
As expected, the exponents
given above for two-dimensional 
SAWs satisfy both these relations.

In three dimensions the most precise estimates we have are from the Monte
Carlo work of one of us~\cite{C10, C13}, with $\mu = 4.684039931
\pm 0.000000027,$
$\gamma= 1.156957 \pm 0.000009$ and $\nu  = 0.587597 \pm 0.000007,$
though the details of the Monte Carlo simulation for the estimation of $\gamma$
have not yet been published. For $\gamma_1$ there are a few estimates in
the literature based on rather short series. Some 30 years ago, Guttmann
and Torrie~\cite{GT84} estimated $\gamma_1 = 0.676 \pm 0.009,$ while the
most recent Monte Carlo estimate is by  Grassberger~\cite{G05} who
estimated $\gamma_1= 0.6786 \pm 0.0012.$ From Barber's scaling relation
this gives $\gamma_{11}=-0.3874 \pm 0.0024.$

There is renewed interest in the values of the exponents $\gamma_1$ and
$\gamma_b$ in this case, as the former exponent arises in recent tests
for the conformal invariance of 3d SAWs~\cite{K13,K15}, and the latter  has
been predicted by the recent scaling argument~\cite{DG15} given above,
and estimation of the exponent $\gamma_b$ would be a useful test of that
scaling relation. More precisely, Kennedy discusses the hitting density
of a SAW in a connected domain when conformally mapped by a function $f$
to another domain. The hitting densities with respect to arc length in the
two domains are related by $|f'(z)|^b,$ where $b$ is a critical
exponent, identified by scaling arguments as 
\begin{align}
b &= \frac{\gamma-2\gamma_1}{2\nu}+\frac{d}{2},\\
&= -\frac{\gamma_b}{2\nu} + \frac{d}{2}.
\label{eq:bscaling}
\end{align}
Note that the above exponent estimates give the prediction $b = 1.3296
\pm 0.0020,$ compared to Kennedy's direct Monte Carlo estimate $b = 1.3303 \pm
0.0003$~\cite{K15}. For the bridge exponent the prediction from the pre-existing
exponent estimates is $\gamma_b = 0.2002 \pm 0.0024.$

In this paper we will estimate critical exponents and amplitudes
associated with bridges and TAWs via series and Monte Carlo methods.

\subsection{Universal amplitude ratio}
Combining the two scaling relations, (\ref{eq:scalinga}) and
(\ref{eq:scalingb}), given in
Sec.~\ref{Intro}, we
obtain
\begin{align}
2\gamma_1 &= \gamma+\gamma_b.
\label{eq:scaling}
\end{align}
This exact relation suggests that the
relationship between amplitudes corresponding to the LHS and RHS of the
equation must be lattice independent, which
in turn leads to a universal amplitude ratio.

One interesting aspect of this ratio is that the objects involved are bridges and
TAWs, which interact with a surface. Loosely speaking, the scaling relation
(\ref{eq:scaling}) may be interpreted as saying that the 
exponent associated with the number of pairs of TAWs of length $n$ (which together have 2 free
ends, and two ends anchored to a surface) is the same as the exponent
associated with the number of pairs made up of a SAW and a bridge, each of length $n$.

In
order for the lattice specific effects to cancel 
it is crucial that the boundary condition where the walk terminates on
the surface is the same for the TAWs and both ends of the
bridges. 

Now, the TAWs are anchored to the surface, and must satisfy the rule that the
entire walk lies above this surface. In contrast, for convenience the
bridges are defined to be asymmetric, with an additional step
prepended to one end to lift it completely out of the surface at one end.

Thus we need modify the definition of a bridge to make it 
symmetric between its two ends. The most natural way to do this is to 
remove
the trivial edge from the start of the bridge. Conveniently, 
a straight-forward bijection exists between 
modified bridges of length $n$ which satisfy
$z^{(0)} \leq z^{(i)} \leq z^{(n)}$ for any $0 \leq i \leq n$,
and the usual bridges of length $n+1$ which satisfy
$z^{(0)} < z^{(i)} \leq z^{(n+1)}$ for any $0 < i \leq n+1$. Hence the
number of modified bridges of length $n$ is just 
$b_{n+1}$, and this is the coefficient that must be used in the
definition of the amplitude ratio.

At the coefficient level, the above considerations lead to the following
sequence of ratios
\begin{align}
K_n &= \frac{t_n^2}{c_n \cdot b_{n+1}} \sim \frac{H^2}{\mu A B}
    n^{2\gamma_1 - \gamma_b - \gamma} \left(1 + {\rm o}(1))\right),\\
K &\equiv \lim_{n \to \infty} K_n  = \frac{H^2}{\mu A B}
\end{align}
where the amplitudes $A$, $B$, and $H$ are from
(\ref{eq:saw_asympt})--(\ref{eq:t_asympt}), and $K$ is a universal constant. 
That is to say, we expect this ratio $K$ to depend only on the spatial
dimension of the lattice.

In terms of generating functions, the amplitude ratio is given as 
the limit 
\begin{align}
\lim_{x \rightarrow x_c^-} \frac{T(x)^2}{\mu \cdot C(x)\cdot B(x)} &= {\rm Universal\,\,\,
constant.}
\end{align}

\subsection{Confidence intervals}
We wish to highlight that the error estimates for the quantities
calculated in this paper are not purely statistical, and hence there is
a degree of subjectivity in obtaining them. In the analysis of
relatively short series, which is the best that can be achieved for non-trivial 3d models
such as SAWs, TAWs, and bridges, the interpretation of the results from
series analysis can be quite subtle, and it is easy to be overly
optimistic in interpreting the convergence (or lack thereof) of a
sequence of improved estimates. For our Monte Carlo results we are
approaching the large $n$ limit, and consequently there is a lesser (but
non-zero) amount of
subjectivity involved.

We have included plots where relevant to allow readers to judge for
themselves how reliable our confidence intervals are.
Loosely, one may think of the confidence intervals given in this paper
as being roughly equivalent to a single standard deviation.

\subsection{Outline}
\label{sec:outline}
In Sec.~\ref{sec:series} we describe the generation of the series data for
simple cubic lattice TAWs and bridges, which is then analysed
in the Sec.~\ref{sec:seriesanalysis}. Thereafter we describe the Monte Carlo
computer experiment in Sec.~\ref{sec:mcmethod}, and analyse the resulting data in
Sec.~\ref{sec:mcanalysis}.
Finally, we discuss possible extensions and
conclude in Sec.~\ref{sec:discussion}.

\section{Generation of series data}
\label{sec:series}

We directly enumerated bridges and TAWs via a straight-forward optimized backtracking
algorithm, whose running time was roughly proportional to the
number of items being computed.

To enumerate TAWs up to a certain length a recursive function
was used that took the set of sites currently visited, the coordinates
of the current site, and the maximum number of bonds left. The function
first adds one to the tally of TAWs of the current length. It
then, assuming there is at least one bond left, chooses each unused
adjacent site (not allowing negative $z$ coordinates) and calls the
function again with that site as the current site, and one fewer sites
remaining.

This algorithm clearly has the function called exactly once for each
TAW enumerated. As each iteration of the recursive function
contains a small amount of code, with only one loop over adjacent sites
(maximum 6), its execution time is exactly proportional to the number of
TAWs enumerated.

Enumerating bridges uses a very similar algorithm. In this case the
recursive algorithm keeps track of an extra variable; the maximum $z$
value reached ($Z_{\rm max}$).  Now instead of adding one to the 
TAW tally on each function call, a check is performed to see if the $z$
coordinate of the current site equals $Z_{\rm max}$.

As described, this algorithm would have execution time proportional to
the number of TAWs, as exactly the same function calls would
be made as for the TAWs. The performance can be improved to
be proportional to the (significantly smaller) number of bridges by
pruning the search tree when  the current TAW cannot ever
result in a valid bridge. This can be prohibitively expensive to compute
exactly; a fast and very effective heuristic is to prune any branches
when the $z$ coordinate of the current point plus the number of
remaining bonds is less than $Z_{\rm max}$. This is somewhat conservative;
it will not prune every dead branch as there might be an occupied site
in the way; however in practice it works very well resulting in a
runtime basically proportional to the number of bridges.

We can reduce the work required to determine $b_n$ by enumerating
irreducible bridges instead, which are the set of bridges that cannot be 
decomposed into a sequence of
more than one bridge. 
Irreducible bridges can also be characterised as the set of bridges which
have at least 3 bonds between any two layers, other than the first
layer,
which must have precisely one bond.
Each bridge can in fact be
uniquely decomposed into a non-empty sequence of irreducible bridges,
and so the
generating function for bridges, $B(x)$, may be expressed in terms of
the generating
function for irreducible bridges, $I(x)$, as
\begin{align}
    B(x) &= \frac{I(x)}{1-I(x)}.
\label{eq:irreducible}
\end{align}
Using this relation one may obtain $b_n$ from the series for the less
numerous irreducible bridges. 

Enumerating irreducible bridges requires a slight modification to the bridges
algorithm. In this case yet another variable $Z_{\rm irr}$ is kept track of
in the recursive function: $Z_{\rm irr}$ is the largest $z$ value below which the
TAW is irreducible (it has at least three bonds in the $z$
direction at each layer). To see if it is a valid irreducible
bridge one now checks that the current $z$ value, $Z_{\rm max}$, and $Z_{\rm irr}$
are all equal. $Z_{\rm irr}$ is easy to keep track of; $Z_{\rm irr}$ is
incremented whenever a vertical step is made upwards from $z=Z_{\rm irr}$,
as long as $Z_{\rm irr}<Z_{\rm max}$.

As described, this algorithm would have the same execution time as the
bridges algorithm. Again, pruning can improve the speed to close to the
significantly smaller number of irreducible bridges. The following
conservative heuristic is used:
\begin{itemize}
\item
If $z \leq Z_{\rm irr} \leq Z_{\rm max}$, prune if the number of remaining steps
is less than $Z_{\rm max}-z$  (this is the same as the bridges pruning);
\item
If $Z_{\rm irr} < z \leq Z_{\rm max}$, prune if the number of remaining steps is
less than $Z_{\rm max}-z+1+2 (z-Z_{\rm irr}) $  (this is due to the need to go
back down to $z=Z_{\rm irr}$, take one step sideways, and then return to
$z=Z_{\rm max}$).
\end{itemize}
We find that his heuristic works very well in practice, and so 
the running time is roughly proportional to the number of
irreducible bridges.

There were a variety of other optimizations used to give constant
improvements, such as assigning each conceivably reachable lattice point
an integer index and pre-computing adjacencies so that points were
represented by integers rather than a tuple of integers.

Symmetries were taken into account by enumerating all prefixes up to
length roughly 8 (it varied somewhat depending on what computer the
enumeration was run on), canonicalizing them to account for symmetries,
and assigning to equivalence classes. Then the recursive algorithm
described above was used on each equivalence class, and the results
multiplied by the number of members of the equivalence class.
The enumerations for each equivalence class were summed for the final
result.

It was then straight-forward to make the algorithm parallel (to run on
many different processors simultaneously). Each equivalence class can be
computed independently, and so were parcelled out to multiple
processors, with the only interaction being the final summation.

The algorithm as described above is in no way specific to three-dimensional
systems, and indeed
was made to also work in other dimensions. Different lattice types could
also be used, but this may affect the heuristics used for pruning dead
branches of the search tree.

The memory use is insignificant; execution time is the constraint.

\section{Analysis of series}
\label{sec:seriesanalysis}

As usual, we assume that the critical behaviour of TAWs  is
given by
\begin{align}
T(x) &= \sum_{n \ge 0} t_n \cdot x^n \sim  A(1 - \mu \cdot
x)^{-\gamma_1}(1 + A'(1 - \mu \cdot x)^{{\Delta_1}}),
\end{align}
and that of bridges by
\begin{align}
B(x) &= \sum_{n \ge 0} b_n \cdot x^n \sim C(1 - \mu \cdot
x)^{-\gamma_b}(1 + C'(1 - \mu \cdot x)^{{\Delta_1}}).
\end{align}
Here $t_n$ ($b_n$) is the number of $n$-step TAWs  (bridges),
counted up to translations. $\mu$ is the growth constant, for which we
assume Clisby's~\cite{C13} estimate $\mu =4.684039931 \pm 0.000000027.$
Here ${\Delta_1} = 0.528 \pm 0.012$~\cite{C10} is the leading 
correction-to-scaling exponent.  
$A,$ $A',$ $C$ and $C'$ are critical
amplitudes (N.B., these are simply related to, but not the same as, the
amplitudes for the series defined previously). 

In analysing the data, we have of course assumed the existence of the
various critical exponents (which has not been proved), and have also
used the best available estimate of the growth constant, given above,
and the estimate of the correction-to-scaling exponent ${\Delta_1}.$
There are a number of estimates in the literature for this exponent,
obtained by a variety of methods, including field theoretical methods,
series methods and Monte Carlo methods. Estimates range from 0.47 up to
0.57, with non-overlapping error bars. The recent Monte Carlo work of
one of us~\cite{C10} gives $\Delta_1 = 0.528 \pm 0.012,$ and it is this value that
we will use in our Monte Carlo analysis.  However our series work is
less precise than the Monte Carlo work, and for that purpose taking
${\Delta_1} = 1/2$ is sufficiently precise.

In Appendix~\ref{sec:enumerations} we give in Table \ref{tab:one} the series for 
TAWs up to length 26, and for bridges up to length 28. The bridge series
is longer, as we actually counted the far less numerous 
irreducible bridges, and reconstructed the bridge generating function
from the irreducible bridge generating function. We also recorded the
heights of the irreducible bridges, and the full two-variable generating
function for irreducible bridges, counting both length and height, is
given in Table \ref{tab:two} in the appendix. 
The relationship between bridges and irreducible bridges is given in
(\ref{eq:irreducible}).

We first analysed both the TAW and bridge series by using the standard
method of differential approximants~\cite{G89}. We initially attempted
unbiased analyses of both the TAW series and the bridge series. The
results were disappointingly imprecise, despite choosing to use third
order differential approximants, which should, in principle, be able to
accommodate the expected confluent singularity structure.

We had always intended to use biased differential approximants, as these
are expected to give much more precise estimates of the exponents. But
the disappointing lack of well-converged estimates in the unbiased case
warns us not to be too hopeful of obtaining excellent results in the
biased case. Despite this, the estimates appeared tolerably well
converged, and for TAWs we estimated $\gamma_1 = 0.683 \pm 0.005,$ while
for bridges we estimated $\gamma_b = 0.199 \pm 0.004.$ The estimate of
$\gamma_1$  just overlaps the earlier Monte Carlo estimate of Grassberger
\cite{G05}, and also just overlaps our much more precise Monte Carlo estimate
quoted below. We do not know why this estimate of $\gamma_1$ is not more
accurate. It is possible that the amplitude of the correction-to-scaling
exponent is quite large, but we were unable to confirm this from the
available data. The corresponding estimate of the bridge exponent
$\gamma_b$ is in excellent agreement with our Monte Carlo estimate, given below,
though that estimate is again significantly more precise. We do not give
more details or tables of data here, as in the next subsection we give a
more precise analysis.

A completely different method of analysis was also tried. We first
calculated the ratios of alternate coefficients, $r_n = \sqrt{
\frac{t_n}{t_{n-2}} },$ where we have used alternate coefficients to
minimise the effect of the anti-ferromagnetic singularity located at
$x=-1/\mu.$ Then by standard ratio method techniques~\cite{G89}, one
expects the ratios to behave as 
\begin{align}
\label{eq:grn}
r_n \sim \mu\left(1 + \frac{g}{n} + \frac{c}{n^{1+{\Delta_1}}}+
    \frac{d}{n^2} + \cdots\right).
\end{align}
Here, for TAWs $g = \gamma_1-1,$ and $c$ and $d$ are
constants, while for bridges $g=\gamma_b-1.$ Using the estimate of $\mu$ given above, we can then estimate
the exponent $g$ as the limit of the sequence $g_n$ defined by
\begin{align}\label{eq:gn}
g_n &= \left ( \frac{r_n}{\mu}-1 \right )n \sim g + \frac{\rm const.}{n^{\Delta_1}} + \frac{\rm const.}{n} + \cdots .
\end{align}

We used the inbuilt fitting function of Maple to fit the elements of the
sequence $g_n$ from $n=n_{min}$ to $n=n_{\rm max},$ where $n_{\rm max}=28,$ to
\begin{align}
g_n &=  g + \frac{\rm const.}{n^{\Delta_1}} +\frac{\rm const.}{n},
\end{align}
with $n_{min}$
varying from $10$ up to $21.$ The estimates of $g_n$ were reasonably
well converged, and in this way we estimated $g=-0.323 \pm 0.003,$ or
$\gamma_1 = 0.677 \pm 0.003.$ A similar analysis for the bridge series
was less well converged. Fitting just to
\begin{align}
g_n &= \left( \frac{r_n}{\mu}-1 \right) n \sim g + \frac{\rm const.}{n^{\Delta_1}}
\end{align}
gave a
decreasing sequence of estimates of $g_{\rm bridges},$ while fitting to
\begin{align}
g_n &=  g + \frac{\rm const.}{n^{\Delta_1}} +\frac{\rm const.}{n}
\end{align}
as done for
TAWs gave an increasing sequence of estimates of
$g_{\rm bridges}.$ Averaging corresponding entries gave a fairly stable
sequence, from which we estimated $g_{\rm bridges} = -0.802 \pm 0.005,$ or
$\gamma_b = 0.198 \pm 0.005.$

Combining our results from the two methods of analysis, we give as our series estimates from this data, $\gamma_1 = 0.680 \pm 0.003$ and $\gamma_b =
0.1985 \pm 0.004.$ Again, we do not give more details or tables of data here, as in the next subsection we give a more precise analysis.

\subsection{Series extension and subsequent analysis}
\label{sec:seriesextension}

The analysis discussed above is based on the exact series coefficients
given in Appendix~\ref{sec:enumerations}.
We can however obtain accurate approximations to
the next five to seven terms of the series, which can then be used in
our ratio analysis. Because every differential approximant that uses all
the available series coefficients implicitly predicts all subsequent
coefficients, we realised that we could utilise this observation to
calculate, approximately, all subsequent coefficients. Of course the
number of significant digits decreases as the number of predicted
coefficients increases, but as we show below, we can get useful
estimates of the next six or seven coefficients. 

For every approximant using all the known coefficients, we generated the
subsequent six or seven coefficients. We observe that the predicted
coefficients agree to a certain number of significant digits among all
the approximants, and we take these as our estimates. That is to say,
assume we know the coefficients $a_n$ for $n \in [0,N_{\rm max}].$ We then
predict the coefficients
$a_{N_{\rm max}+1},\,a_{N_{\rm max}+2},\,\cdots,a_{N_{\rm max}+7}.$ Our estimate of
each such coefficient is given by the average of the values predicted by
the differential approximants. We reject obvious outliers, which are
infrequent, and quote only those digits for which all approximants
agree. So for example if the first 8 digits agree, and the coefficient
is predicted to be a 20 digit integer, we will quote the coefficient as
the 8 predicted digits followed by 12 zeros. Not surprisingly, we find
the greatest number of digits are predicted for $a_{N_{\rm max}+1},$ with
the number of digits slowly decreasing as we generated further
coefficients.

These predicted coefficients are not suitable for including in
our differential approximant analysis, as they were derived from lower
order differential approximants. However for ratio type analyses they
are very well suited, as discrepancies in say the seventh or eighth
significant digits will not affect the ratio analysis in the slightest.
This is particularly useful in those situations where we suspect there
might be a turning point in the behaviour of ratios or their
extrapolants with our exact coefficients, as these approximate
coefficients are more than accurate enough to reveal such behaviour, if
it is present.

As a demonstration of this method, assume we only have 23 terms in the
bridge generating function, and we'll predict the next 6 coefficients.
In Table \ref{tab:bridge-pred} we show the predicted and exact
coefficients. (Note that the coefficient of $x^{29}$ is not exact, but
is as predicted from a longer series, as discussed below).

\begin{table}[htbp]
\begin{center}
\begin{tabular}{p{30pt} p{110pt} p{110pt} }
\hline
$n$ &Predicted bridges &  Actual bridges \\
\hline
24 &	$7.115933 \times 10^{14}$  &  711593257794069\\
25 &	$3.235637 \times 10^{15}$ & 3235634079777801 \\
26   & $1.472844 \times 10^{16} $&   14728414578753489 \\
27   &    $6.71105\times 10^{16}$                                &   67110197685388181 \\
28   &  $3.06077\times 10^{17}$                                  &   306074586987649389 \\
29 & $1.39718 \times  10^{18} $& $1.397156394  \times 10^{18}$\\
\hline
\end{tabular}
\caption{Approximate, predicted coefficients of simple cubic lattice  bridges of length $n,$ compared to exact values.}
\label{tab:bridge-pred}
\end{center}
\end{table}

Note that in every case the predicted coefficient agrees with the exact
coefficient to the order stated, with an uncertainty of a few parts
in the last quoted digit. This demonstrates the utility of the method.

In Table~\ref{tab:one-pred} are the predicted coefficients from the full series, where, as
explained above, we expect errors to be confined to the last quoted
non-zero digit. (Of course it is possible that the last digit we are
confident of is 0, but nothing is lost in glossing over this).

\begin{table}[htbp]
\begin{center}
\begin{tabular}{p{30pt} p{110pt} p{110pt} }
\hline
$n$ &TAWs & Bridges \\
\hline
27 & $6.99275590 \times 10^{17}$   & {\rm exact} \\
28 &	$3.2418211 \times 10^{18}$ & {\rm exact} \\
29 & $	1.5038283\times 10^{19}$   & $ 1.397156394 \times 10^{18}$\\
30 &	$6.9761657 \times 10^{19}$ & $ 6.38289508 \times 10^{18}$\\
31 &	$3.237937 \times 10^{20}$  & $ 2.91825221  \times 10^{19}$\\
32 &	$1.502907 \times 10^{21}$  & $ 1.33518243  \times 10^{20}$\\
33   & $6.979108 \times 10^{21}$   & $ 6.1129712  \times 10^{20}$\\
34   &                             & $ 2.8005385 \times 10^{21}$\\
35   &                             & $ 1.2837858 \times 10^{22}$\\
\hline
\end{tabular}
\caption{Approximate, predicted coefficients of simple cubic lattice TAWs and bridges of length
$n$.}
\label{tab:one-pred}
\end{center}
\end{table}

\begin{figure}[!ht] 
\centering
\includegraphics[width=0.6\textwidth]{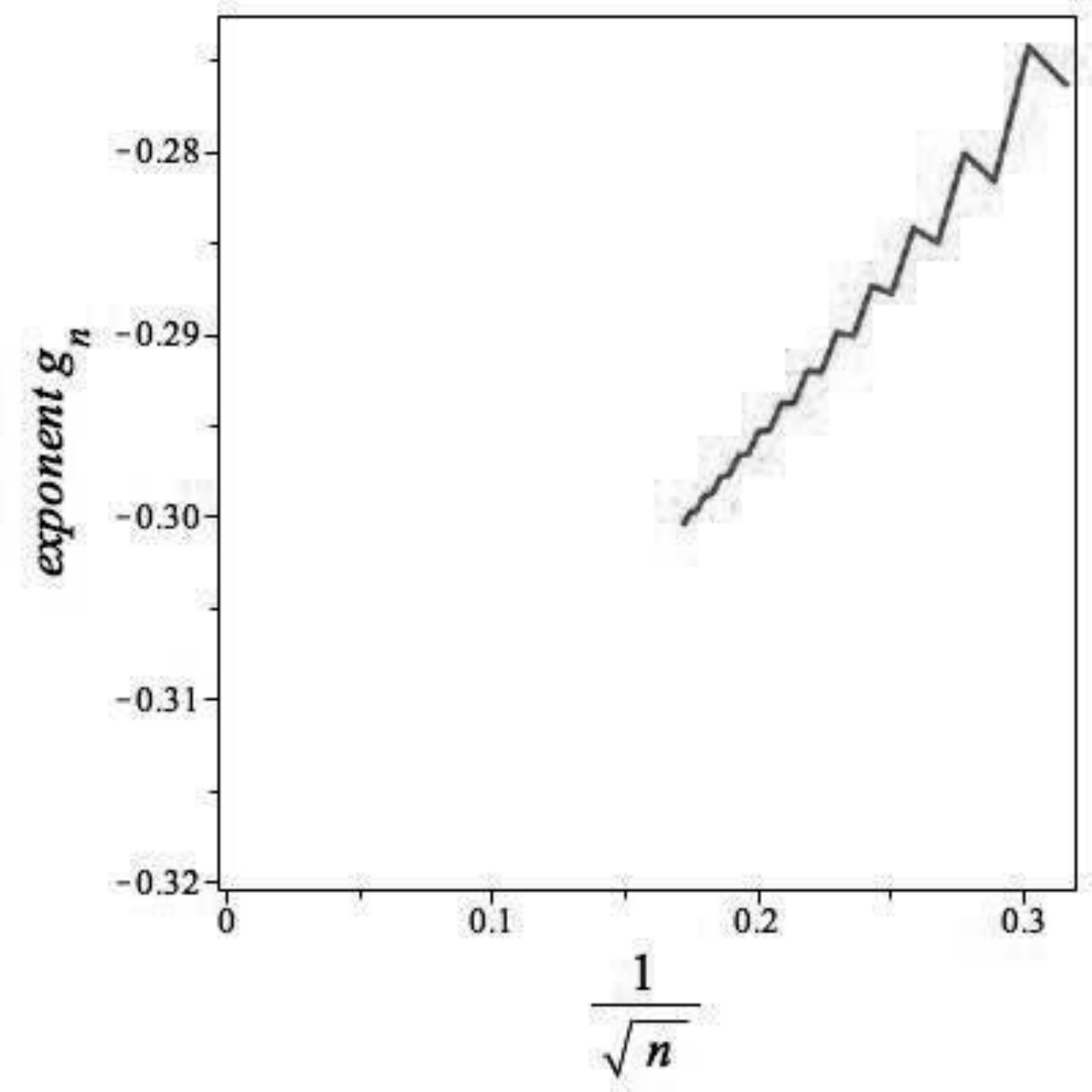} 
\caption{Exponent sequence $g_n$ for TAWs}
\label{fig:g-taws}
\end{figure}

We repeated the basic ratio analysis described in the previous section,
and show in Fig.~\ref{fig:g-taws} the estimators $g_n$ of the exponent
$g$ for TAWs plotted against $1/\sqrt{n},$ as is appropriate, see
(\ref{eq:gn}). Strong oscillations can be seen, indicating the presence
of a significant anti-ferromagnetic term.

\begin{figure}[ht] 
\centering
\includegraphics[width=0.6\textwidth]{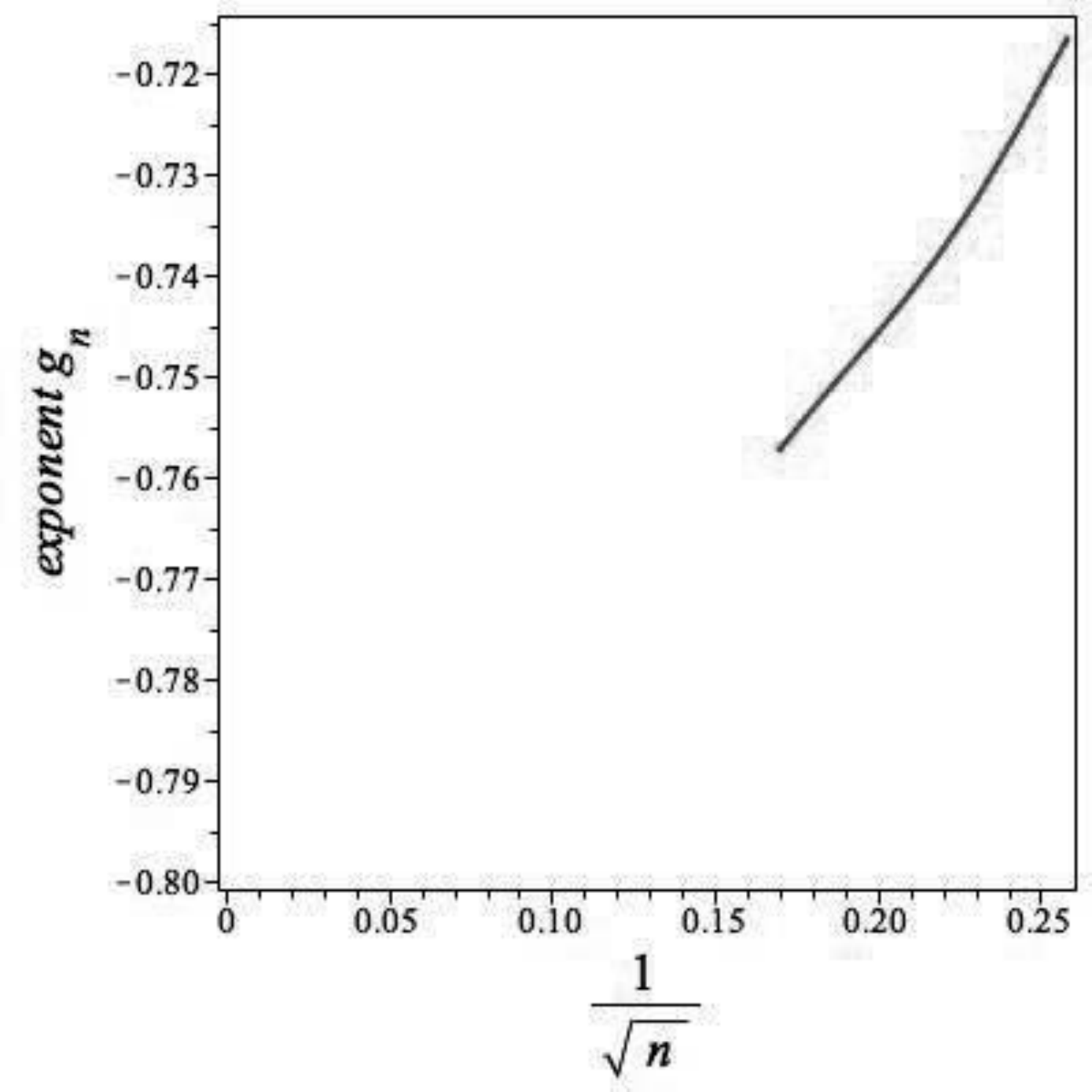} 
\caption{Exponent sequence $g_n$ for bridges}
\label{fig:g-bridges}
\end{figure}

We fitted the data using the curve-fitting features of Maple, simply by
fitting the data points $g_n$ from 12 up to $N_{\rm max}-k$ for $k=0 \ldots
9,$ where $N_{\rm max} = 35$ in the case of bridges and is 34 in the case of
TAWs, assuming $g_n$ is quadratic in $1/\sqrt{n},$ as implied by 
(\ref{eq:gn}). The sequence elements $g_n$ are shown in Table
\ref{tab:gn} below.
\begin{table}[htbp]
\centering
\begin{tabular}{p{30pt} p{80pt} p{80pt} }
\hline
$k$ &$g_n$ TAWs & $g_n$ bridges \\
\hline
9 & -0.34067 & -0.86711\\
8 & -0.34120 & -0.85167 \\
7&-0.33687 & -0.83958 \\
6&-0.33707 & -0.83011 \\
5& -0.33429 & -0.82278 \\
4& -0.33432 & -0.81715 \\
3& -0.33244& -0.81287 \\
2& -0.33240 & -0.80970 \\
1& -0.33207 & -0.80739 \\
0 & -0.33098 & -0.80577 \\
\hline
Limit & $-0.324 \pm 0.002$ & $-0.801 \pm 0.002$ \\
\hline
\end{tabular}
\caption{Sequences of exponent values $g_n$ for TAWs and bridges.}
\label{tab:gn}
\end{table}

These sequences have been simply extrapolated, using every alternate
term in the case of TAWs, to give the limits below. Note that the extra,
approximate, terms are (a) quite precise enough for this sort of
analysis, and (b) move the data into a regime where the extrapolations
are much clearer than those using only the exact coefficients. The
extrapolation method was based on the observation that the numerical
gaps between successive (or alternate, as appropriate) coefficients
apparently decrease as a geometric sequence, so this was summed.

As a result of this analysis we find $\gamma_{1} = 0.676 \pm 0.002,$ and
$\gamma_b = 0.199 \pm 0.002.$ These are consistent with, but more
precise than, the results obtained in the previous section, using only
the exactly known coefficients. In subsequent sections we discuss the
Monte Carlo results, which are substantially more precise than the
series results. However, as we show below, there is one property which
is accessible from series analysis which is not easy to determine by
Monte Carlo analysis, and that is the behaviour of {\em spanning bridges} in a
slab of given thickness, at the critical temperature, as the slab
thickness increases. We discuss this in Sec.~\ref{sec:theta} below.

\subsection{Universal amplitude ratio \label{sec:uar}}

\begin{figure}[!ht] 
\centering
\includegraphics[width=0.6\textwidth]{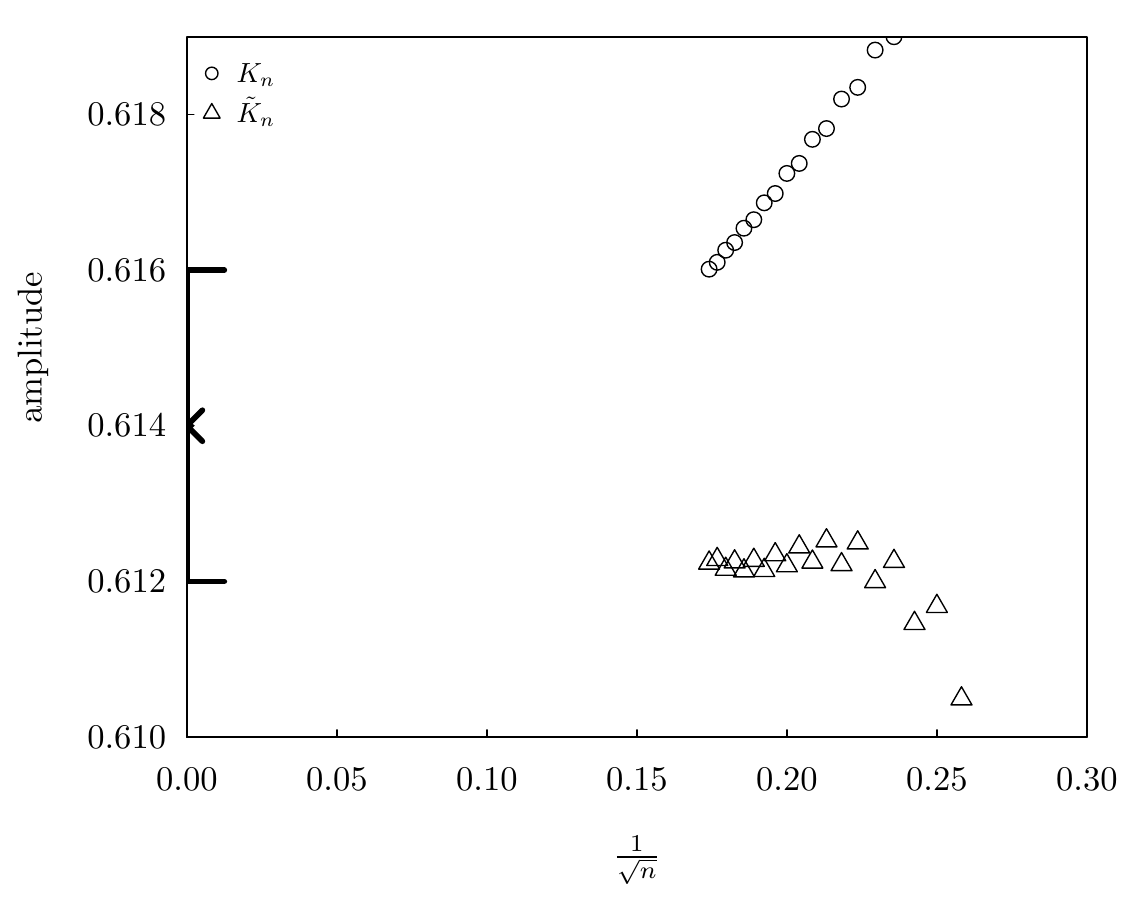} 
\caption{Amplitude estimators ${K}_n$ and $\tilde{K}_n$ for universal
amplitude $K.$ Our best estimate from the series analysis $K = 0.6140
\pm 0.0020$ is also shown.
}
\label{fig:K-amp}
\end{figure}

From our series, and known series for 3d SAWs~\cite{CLS07,SBB11}, we have
calculated $K_n$ for $n \le 27$ using exact coefficients, and for $n \le
34$ using the last seven approximate coefficients.  From the 
asymptotic form of the coefficients, we expect the asymptotic form of
the amplitude to be 
\begin{align}
K_n &\sim K\left (1 +
\frac{k_1}{n^{\Delta_1}}+\frac{k_2}{n} + \cdots \right).
\end{align}
Accordingly, we extrapolated
$K_n$ against $1/n^{\Delta_1},$ using our approximate value $\Delta_1 =
1/2.$ As can be seen in Fig.~\ref{fig:K-amp}, the sequence $K_n$ is monotone decreasing, 
as far as we have data,
with the last entry being $K_{34}=0.61601\ldots,$ which suggests that
this is an upper bound on $K.$ The ratio plots against $1/\sqrt{n}$
exhibit significant curvature, consequently we could only form the
estimate $K > 0.609. $ Combining these numerical confidence bounds, our
first estimate is $K = 0.6125 \pm 0.0035.$ We also extrapolated $K_n$
against $1/n,$ and those plots exhibited similar, though reduced
curvature.
The increased
linearity when plotting against $1/n$ suggests that the amplitude of the
leading correction-to-scaling term is rather weak, and that the sub-dominant
behaviour for $n \leq 34$ is largely given by the O$(1/n)$ term. Accordingly, we
eliminate this term by forming the estimators
\begin{align}
\tilde{K}_n &= \frac{n\cdot K_n - (n-2)\cdot K_{n-2}}{2}.
\end{align}
Note that alternate terms are used to reduce oscillations caused by the
anti-ferromagnetic singularity present in all three generating functions
at $x=-1/\mu$.
We expect 
\begin{align}
\tilde{K}_n &\sim K\left (1+\frac{k_1}{n^{\Delta_1}}+ {\rm
o}\left(\frac{1}{n}\right) \right ).
\end{align}
We show in Fig.~\ref{fig:K-amp} a plot of
$\tilde{K}_n$ against $1/\sqrt{n},$
and the importance of the extra estimated terms is now seen, as we are
just getting the first clear indication of a turning-point in our
estimates. If we plot the odd and even terms separately, this is even
clearer. From this figure, and the separate odd-term and even-term plots
(not shown), we can give the refined estimate $K = 0.6140 \pm 0.0020,$
which is shown on the graph.

A similar analysis for the square lattice was also conducted. There we
have much longer series available, and the dominant correction is
O$(1/n)$, as in two dimensions ${\Delta_1} =1.5.$ In that case we obtained
$K_{2d} \approx 0.57280$ where we expect the error to be confined to
the last digit.

\subsection{Behaviour as strip width grows.}
\label{sec:theta}

The data we have generated also allows us to estimate the exponent
characterising the behaviour of the generating function of bridges, at
the critical point, which span a strip of width $T$ as $T \to \infty.$
For two-dimensional SAWs, under the assumption that these are
describable in the scaling limit by ${\rm SLE}_{8/3},$ one has~\cite{LSW04}
\begin{align}
B_T(x_c) &\sim \frac{\rm const.}{T^{1/4}}.
\end{align}
For the simple cubic
lattice, we expect similar behaviour, so that
\begin{align}
B_T(x_c) &\sim \frac{\rm const.}{T^{\theta}}.
\end{align}
From the data for simple cubic bridges,
given in the appendix, we can construct the first 28 terms in the
generating function for bridges in a strip of width $T.$ Note that such
generating functions have $x^T$ as their first non-zero term, as one
needs at least $T$-steps to span a strip of width $T.$ Note too that the
generating functions will diverge at a critical value $x_T > x_c,$ where
$x_c = 1/\mu$ is the bulk critical value. So evaluating $B_T(x_c)$ means
we are evaluating these ordinary generating functions at a value inside
their radius of convergence.

In order to do this as effectively as possible, we have used Pad\'e
approximants. For each width $T,$ we have constructed a number of
Pad\'e approximants utilising all, or nearly all, of the coefficients,
and evaluated these at $x_c.$ As $T$ increases, the spread among the
approximants for a given value of $T$ increases, and we found we could
only get useful estimates for $T \le 9.$ These were $0.6967886$,
$0.5565774$, $0.4720324$, $0.4144$, $0.3728$, $0.3399$, $0.3132$,
$0.2915$, $0.273$
for $T = 1, \ldots, 9.$ We do not quote errors, but expect the error in
each case to be confined to the last quoted digit. A log-log plot of
these data, which should have gradient $ -\theta,$  shows slight
curvature. Accordingly, we calculate the {\em local gradient} given by
\begin{align}
-\theta_T &= \frac{\log{B_T(x_c)}-
\log{B_{T-1}(x_c)}}{\log(T)-\log(T-1)}.
\end{align}
We plot $\theta_T$ against
$T^{-1/2},$ reflecting the correction-to-scaling exponent around $0.5,$
which gives a straight line (with minor oscillations). Extrapolating
this with a straight-edge, which is an appropriate level of
sophistication given the paucity of data we have, gives a best estimate
of $\theta = 0.7.$ If the data is representative of the asymptotic
behaviour, as it appears to be, quoting a confidence interval of $\pm
0.1$ is appropriately conservative.

We can also give an heuristic argument, based on the ideas given in
Beaton et al.~\cite{BGJL15}, as to the expected value of this exponent
in terms of known exponents. We first argue that the measure of bridges
of length $\le n$ should scale as $n^{\gamma_b},$ where $\gamma_b$ is
the bridge exponent introduced above. Now normalised bridges of height $T$
typically have $T^{1/\nu}$ steps. So the measure of bridges of height
$\le T$ behaves asymptotically as $(T^{1/\nu})^{\gamma_b}.$  Then
bridges of height exactly  $T$ should scale as $(T^{\gamma_b/\nu-1})\sim
T^{-\theta},$ i.e.
\begin{align}
\theta &= 1 - \frac{\gamma_b}{\nu}.
\label{eq:thetascaling}
\end{align}

In two dimensions $\nu=3/4,$ and $\gamma_b=9/16,$ so
$\theta=1/4,$ as expected. In three dimensions we have $\nu = 0.587597
\pm 0.000007,$ and our series estimate for $\gamma_b$ is $0.199 \pm 0.002$, so the above
equation gives $\theta = 0.661 \pm 0.004$, which is consistent with the
direct estimate.

\section{Generation of Monte Carlo data}
\label{sec:mcmethod}

For our computer experiments we utilise the pivot algorithm,
a Markov chain Monte Carlo scheme which allows for the
efficient sampling of self-avoiding walks and related objects. 
The pivot algorithm was invented by
Lal~\cite{L69} but first studied in detail by Madras and
Sokal~\cite{MS88}. In particular, we use the recent implementation
of one of us~\cite{C10,C10a}, which improved on earlier
important work by Kennedy~\cite{K02}.

The pivot algorithm is a method to sample self-avoiding
walks and TAWs of fixed length, which naturally gives estimates for
quantities associated with the mean size and shape of a SAW, such as
$\nu$. However, it is less obvious how to
extend the method to calculate quantities which inherently depend on
different lengths. We get around this difficulty by calculating ratios
between different kinds of walks. The basic method is to efficiently
sample walks in some larger set, and then at each time step test whether
the walk belongs to some smaller set. For the sake of efficiency it is
crucial that the ratio of
the sizes of these two sets is not exponentially small; in practice the
ratio is a negative power of the length $n$.

The two quantities we estimate are the probability that a self-avoiding
walk is a TAW, and the probability that a TAW is
a modified bridge which has had its first step removed. 
We choose to estimate this ratio for two reasons: it is more
straight-forward for our implementation of the pivot algorithm to adapt
it in this way\footnote{Technical aside: this is because we only need to
check that the ends of the walk lie on opposite faces of the minimum
bounding box, without needing to check if additional sites might
lie on the faces.}, and also because it facilitates the estimation of
the
amplitude ratio $K$ from Sec.~\ref{sec:uar} that we anticipate is universal.

For the first computer experiment,
we sampled self-avoiding walks via the pivot algorithm, and for each time
step we tested whether the SAW is in fact also a TAW. 
Our observable is the indicator function $\chi_h(\omega)$, which is 1 if
$\omega$ is a TAW, and zero otherwise.
When we take the expectation of $\chi_h$ with respect to the uniform
distribution on the set of SAWs of length $n$,
from (\ref{eq:saw_asympt}) and (\ref{eq:t_asympt}) we have 
\begin{align}
\langle \chi_h \rangle_n &= \frac{t_n}{c_n} \\
&\sim \frac{H \mu^n n^{\gamma_1-1}}{A \mu^n
n^{\gamma-1}} \\
&\sim \frac{H}{A} n^{\gamma_1-\gamma}.
\end{align}
Thus estimating the mean value of this observable for a variety of
lengths
allows us to estimate the \emph{difference} between the 
exponent for TAWs, $\gamma_1$, 
and the exponent for SAWs, $\gamma$.

Similarly, for our other computer experiment we sampled TAWs
via the pivot algorithm (N.B., the pivot algorithm
is ergodic in this case, even though no proof has appeared in the
literature as far as we know), and for each time step we tested whether
the TAW is indeed a bridge with its first step removed.  Our observable is the
indicator function $\chi_b(\omega)$, which is 1 if $\omega$ is a
bridge with its first step removed, and zero otherwise. 
When we take the expectation of $\chi_b$ with respect to the uniform
distribution on the set of TAWs of length $n$
($\mathcal{H}_n$), from (\ref{eq:bridge_asympt}) and (\ref{eq:t_asympt}) we have 
\begin{align}
\langle \chi_b \rangle_{\mathcal{H}_n} &= 
\frac{b_{n+1}}{t_n}\\
&\sim \frac{B \mu^{n+1} (n+1)^{\gamma_b-1}}{H \mu^n n^{\gamma_1-1}} \\
&\sim \frac{B\mu}{H} n^{\gamma_b-\gamma_1}.
\end{align}
This allows us to estimate the \emph{difference} between the bridge
exponent $\gamma_b$ and the TAW exponent $\gamma_1$.

One key detail of these simulations, which is similar to the calculation
of the connective constant $\mu$ in~\cite{C13}, is that if the pivot site is selected
uniformly at random then this would result in an integrated
autocorrelation time for the ratio which is $\Omega(n)$.
The reason that this is true for $\chi_h$, in brief, is that
a positive fraction of SAWs
must have at their end a configuration 
which is of some fixed length $k$ that cannot ever be attached to the
surface. The shortest possible example for the beginning of a self-avoiding
walk which can never be a TAW on the square lattice is shown
in Fig.~\ref{fig:trapped}. Whenever the beginning of a walk assumes such a
configuration, it must necessarily be the case that
$\chi_h(\omega) = 0$.
\begin{figure}[htb]
\begin{center}
\begin{tikzpicture}[ultra thick,scale=0.70]
\root
\sww \u \l \d \d \r \r \ew
\begin{scope}[shift={(1,-1)}]
\sdw \u \r \u \u \r \r \d \l \d \r \r \u \ew
\end{scope}
\end{tikzpicture}
\end{center}
\caption{Minimal walk beginning which can never be a TAW on the square lattice (solid line) with a possible
extension (dashed line). \label{fig:trapped}}
\end{figure}

If we were to sample from the $n$ possible pivot sites uniformly at random, then the probability
of selecting one of the first $k$ sites is $k/n$. Thus we expect
that it must take on average O$(n)$ time steps before a pivot site is
selected sufficiently closely to the end to have any chance in changing
$\chi_h$ from 0 to 1. A similar argument applies to the sampling of
TAWs and the estimation of $\chi_b$.

For each of the computer experiments, 
instead of sampling pivot sites uniformly at random on the $n$-step SAWs
and TAWs we
preferentially sample close to each of the ends, using the following
three distributions for pivot sites $i \in [0,n-1]$, each selected with
probability 1/3:
\begin{itemize}
\item
Integer $i$ sampled uniformly from the interval $[0,n-1]$;
\item
    Real $x$ sampled uniformly from $[0,\log(n+1))$,  $i = \lfloor e^x \rfloor - 1$;
\item
    Real $x$ sampled uniformly from $[0,\log(n+1)])$,  $i = n - \lfloor e^x \rfloor$.
\end{itemize}
This choice of pivot site sampling distribution guarantees that all
length scales with respect to the distance from each end are rapidly
sampled.
We sample uniformly from all symmetries of the simple cubic lattice,
excluding the identity.

SAWs were initialised using the pseudo-dimerisation algorithm described
in~\cite{C10a}, followed by approximately $20n$ successful pivots,
with pivot sites chosen
uniformly at random, to
equilibriate the Markov chain.

The TAWs were also initialised using pseudo-dimerisation, 
but an additional
sequence of pivot moves was performed until the SAW 
was in the half-space.
For further equilibration, we wanted to ensure that there was no bias
due to the surface, and so instead of sampling pivot sites uniformly
we sampled pivot sites $i \in [0,n-1]$ with probability 1/2 from the following two
distributions:
\begin{itemize}
\item
Integer $i$ sampled uniformly from the interval $[0,n-1]$;
\item
Real $x$ sampled uniformly from $[0,\log(n+1))$,  $i = \lfloor e^x \rfloor - 1$.
\end{itemize}
We performed approximately $40n$ successful pivots in this case.

Each computer experiment was run for 21 different values of $n$, 
with $n = 2^k - 1$ for $k=9,10,\cdots,25$,
with additional data for $n$ at $n=723, 1447, 2895, 5791$, as we found
that there was more information available for our estimates from smaller
values of $n$.
Approximately 20000 CPU hours were used in each experiment, divided equally
between the different lengths, on
SunFire X4600M2 machines with 2.3GHz AMD Opteron CPUs.
These data are reported in Appendix~\ref{sec:mcdata}.

In addition, we invested a relatively small amount of computer time
gathering data for shorter lengths. We
do not report these result here, but show them in graphical form below in
Fig.~\ref{fig:h2bs_d3_full_amplitude}
to illustrate strong corrections-to-scaling.

\section{Analysis of Monte Carlo data}
\label{sec:mcanalysis}

For our analysis we used the method of direct fitting.
We fitted our data from Appendix~\ref{sec:mcdata} with the asymptotic
forms for each of the series, using in each case the term with
the leading correction-to-scaling involving $\Delta_1$. We have also
performed
fits of the dominant term only, 
as an
alternative, robust, but less accurate method for determining the values
of critical parameters.
These more robust fits confirm that it is
reasonable to fit the $\Delta_1$ term in each case.

We used weighted least squares to fit all values for $n \geq
n_{\mathrm{min}}$, and plotted the fits against an appropriate power of
$n_{\mathrm{min}}$ which was of the same order as the first
neglected correction-to-scaling term. We only required linear fits, and
used the statistical programming language R as our tool.
We performed fits with  
$n_{\mathrm{min}} \leq 16383$, as statistical noise for larger
$n_{\mathrm{min}}$ rendered these fits useless.
The fitting forms we used are:
\begin{align}
\langle \chi_h \rangle_n &= \frac{t_n}{c_n}
\sim \frac{H}{A} n^{\gamma_1-\gamma}
\left(1 +
\frac{\text{\rm const.}}{n^{\Delta_1}} + {\rm O}\left(\frac{1}{n}\right) \right)
\\
\log \langle \chi_h \rangle_n &\sim (\gamma_1-\gamma) \log (n +\delta_h)
+ \log\frac{H}{A} + \frac{\text{\rm const.}}{(n+\delta_h)^{\Delta_1}} +
{\rm O}\left(\frac{1}{n}\right) \label{eq:log_chi_h}\\
\langle \chi_b \rangle_{\mathcal{H}_n} &= \frac{b_{n+1}}{t_n} 
\sim \frac{B\mu}{H} n^{\gamma_b-\gamma_1} \left(1 +
\frac{\text{\rm const.}}{n^{\Delta_1}} + {\rm O}\left(\frac{1}{n}\right) \right)
\\
\log \langle \chi_b \rangle_{\mathcal{H}_n} &\sim (\gamma_b-\gamma_1) \log (n +\delta_b)
+ \log\frac{B\mu}{H} + \frac{\text{\rm const.}}{(n+\delta_b)^{\Delta_1}} +
{\rm O}\left(\frac{1}{n}\right) \label{eq:log_chi_b}
\\
\frac{\langle \chi_h \rangle_n}{\langle \chi_b \rangle_{\mathcal{H}_n}}
&= \frac{t_n^2}{b_{n+1} c_n} \sim K n^{2\gamma_1 -
\gamma_b - \gamma} \left(1 +
\frac{\text{\rm const.}}{n^{\Delta_1}} + {\rm O}\left(\frac{1}{n}\right) 
\right) \label{eq:direct_ratio}
\\
\log\frac{\langle \chi_h \rangle_n}{\langle \chi_b \rangle_{\mathcal{H}_n}}
&\sim (2\gamma_1 - \gamma_b - \gamma) \log n + \log K
+ \frac{\text{\rm const.}}{n^{\Delta_1}} +
{\rm O}\left(\frac{1}{n}\right)\label{eq:log_ratio}
\end{align}
The neglected next-to-leading correction-to-scaling terms have powers $-1$,
$-2\Delta_1$, and $-\Delta_2$, which are all approximately $-1$ (which
is why we write O$(1/n)$ as a shorthand).
We chose not to attempt to fit these next-to-leading corrections, due to
the competition between these terms which makes it
difficult to sensibly interpret any fitting procedure. 
In addition, we neglected the anti-ferromagnetic singularity which we
expect to be small for the values of $n$ considered here.
Fortunately
we have the luxury of being able to study the large
$n$ regime where we expect all of these correction terms to be small compared to the
leading $n^{-\Delta_1}$ correction.

As was the case for the series analysis, we biased the leading
correction-to-scaling exponent $\Delta_1$. However, since our Monte
Carlo estimates
are more accurate, we used the best available estimate $\Delta_1 = 0.528 \pm
0.012$~\cite{C10} rather than the approximate value $\Delta_1 = 0.5$.
We performed fits for $\Delta_1 = 0.516, 0.528, 0.540$,
but only plotted the confidence intervals for the central estimate.
We did this to gain an understanding of the relative contributions of
statistical error and the systematic error due to biasing the fits with
an imprecise value for $\Delta_1$.

\begin{figure}[htb]
\begin{center}
\begin{minipage}{0.45\textwidth}
\begin{center}
\includegraphics[width=1.0\textwidth]{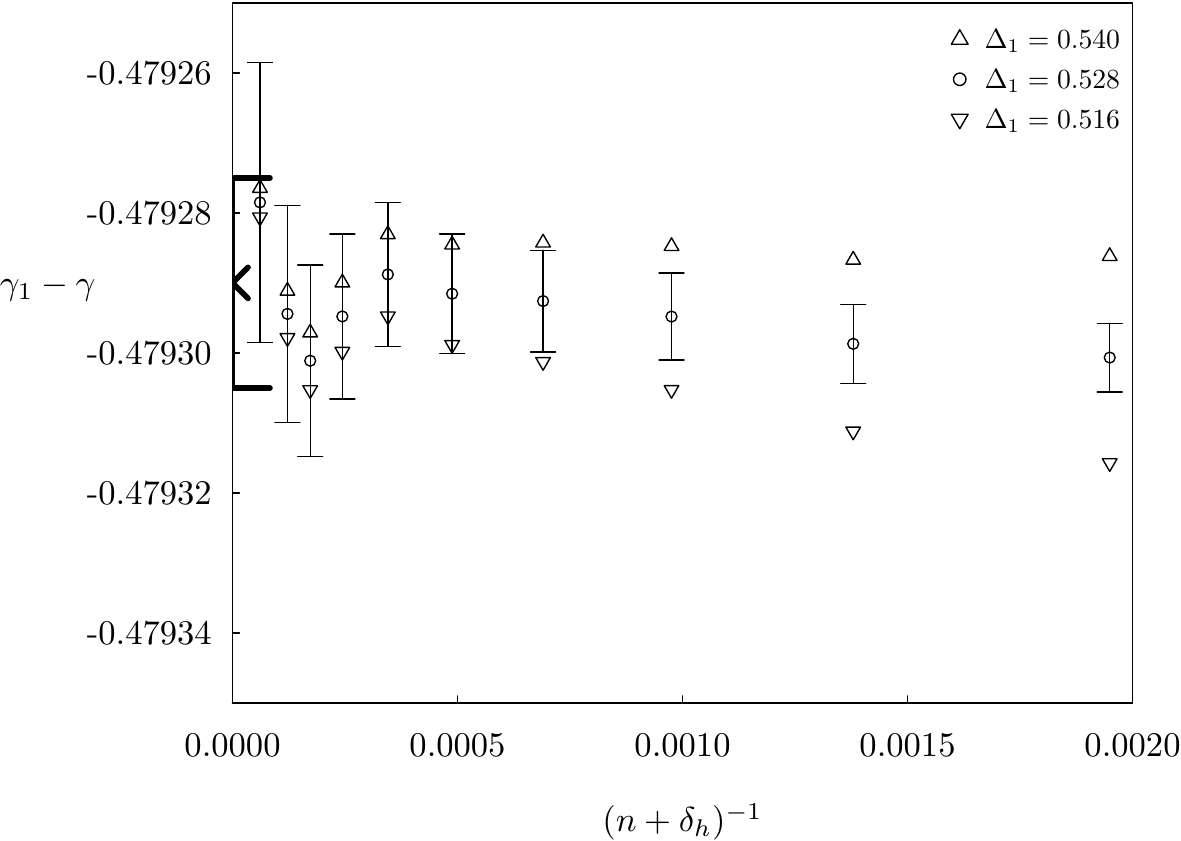}
\end{center}
\vspace{-4ex}
\caption{Estimates of $\gamma_1-\gamma$ from fits to Monte Carlo
estimates of ratios $t_{n}/c_n$, with correction-to-scaling
exponent $\Delta_1 = 0.516, 0.528$, and $0.54$. 
Our extrapolated
estimate, $\gamma_1 - \gamma = -0.479290(15)$, is shown in the plot.
\label{fig:hs_d3_delta}}
\end{minipage}
\hspace{2em}
\begin{minipage}{0.45\textwidth}
\begin{center}
\includegraphics[width=1.0\textwidth]{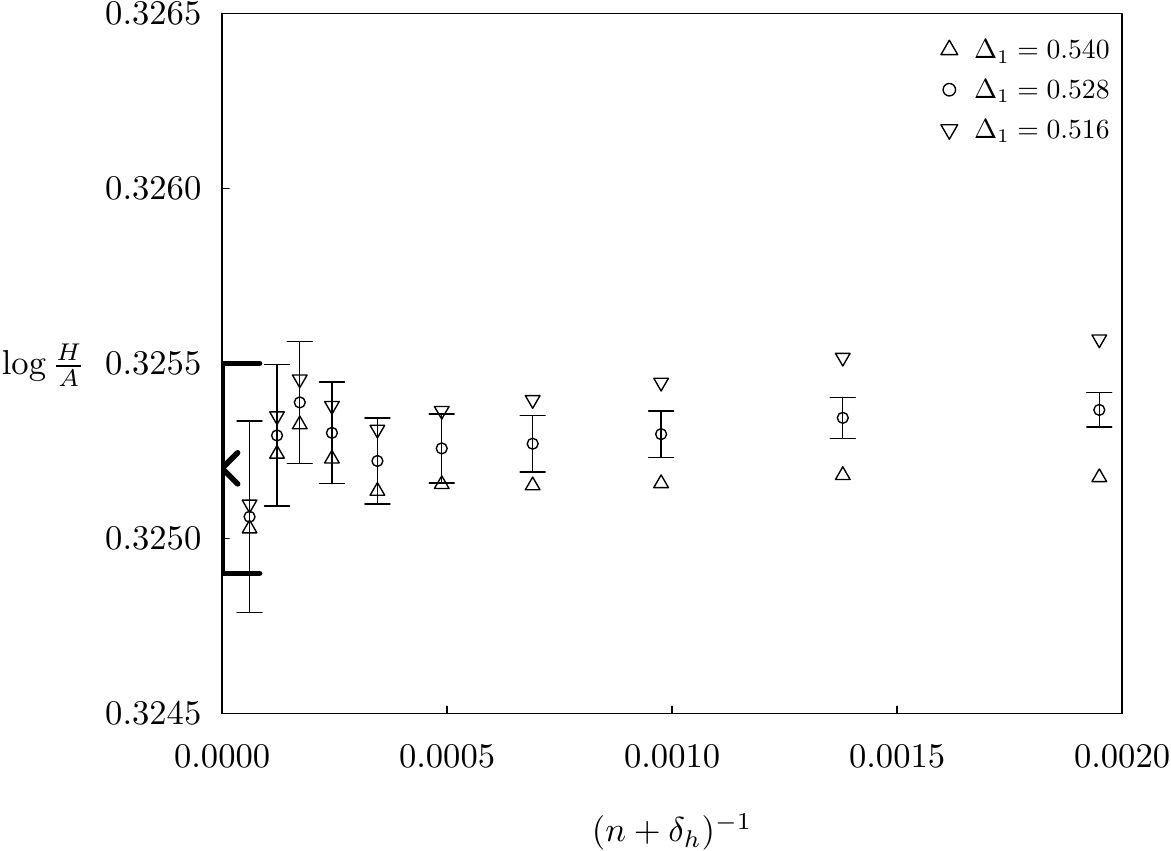}
\end{center}
\vspace{-4ex}
\caption{Estimates of $\log \frac{H}{A}$ from fits to Monte Carlo
estimates of ratios $t_{n}/c_n$, with correction-to-scaling
exponent $\Delta_1 = 0.516, 0.528$, and $0.54$.
Our extrapolated
estimate, $\log \frac{H}{A} = 0.32520(30)$, is shown in the plot. Hence
$\frac{H}{A} = 1.38431(41)$.
\label{fig:hs_d3_delta_amplitude}}
\end{minipage}
\end{center}
\end{figure}

\begin{figure}[htb]
\begin{center}
\begin{minipage}{0.45\textwidth}
\begin{center}
\includegraphics[width=1.0\textwidth]{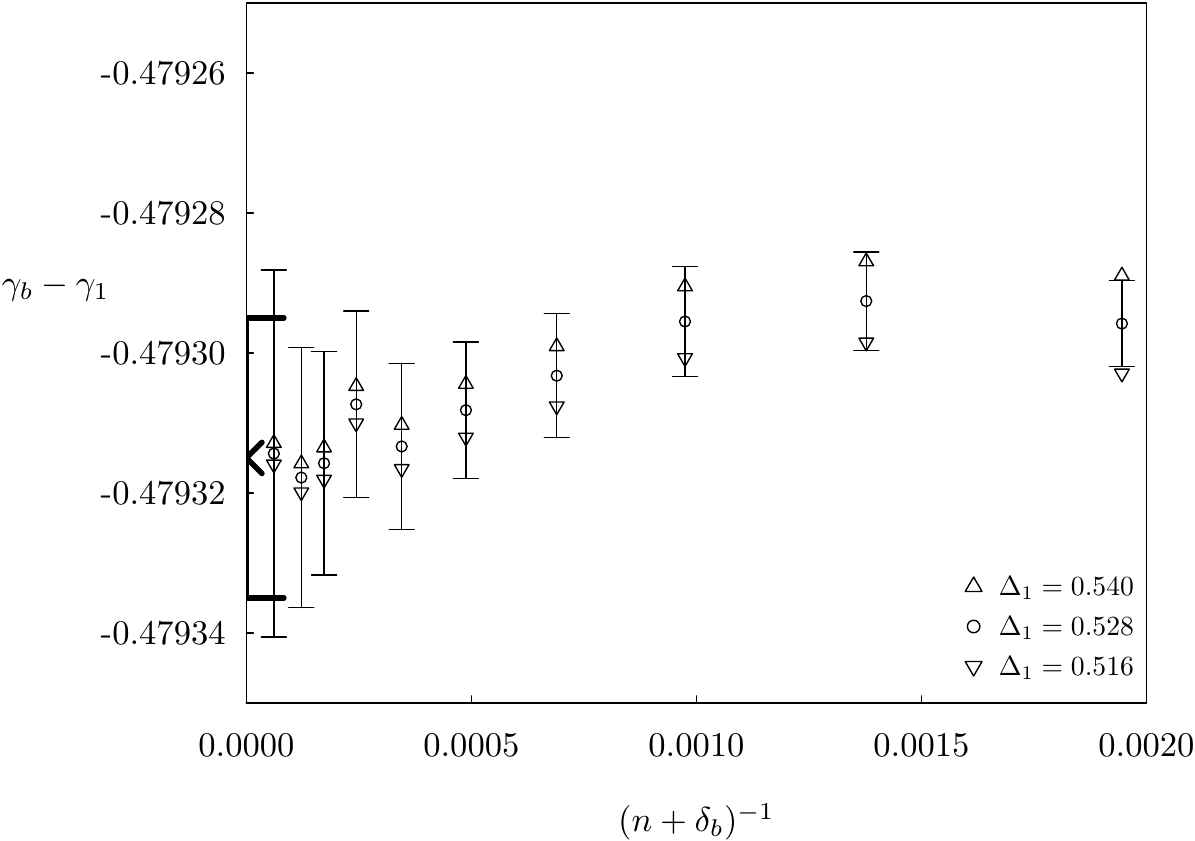}
\end{center}
\vspace{-4ex}
\caption{Estimates of $\gamma_b-\gamma_1$ from fits to Monte Carlo
estimates of ratios $b_{n+1}/t_n$, with correction-to-scaling
exponent $\Delta_1 = 0.516, 0.528$, and $0.54$.
Our extrapolated
estimate, $\gamma_b - \gamma_1 = -0.479315(20)$, is shown in the
plot.
\label{fig:bh_d3_delta}}
\end{minipage}
\hspace{2em}
\begin{minipage}{0.45\textwidth}
\begin{center}
\includegraphics[width=1.0\textwidth]{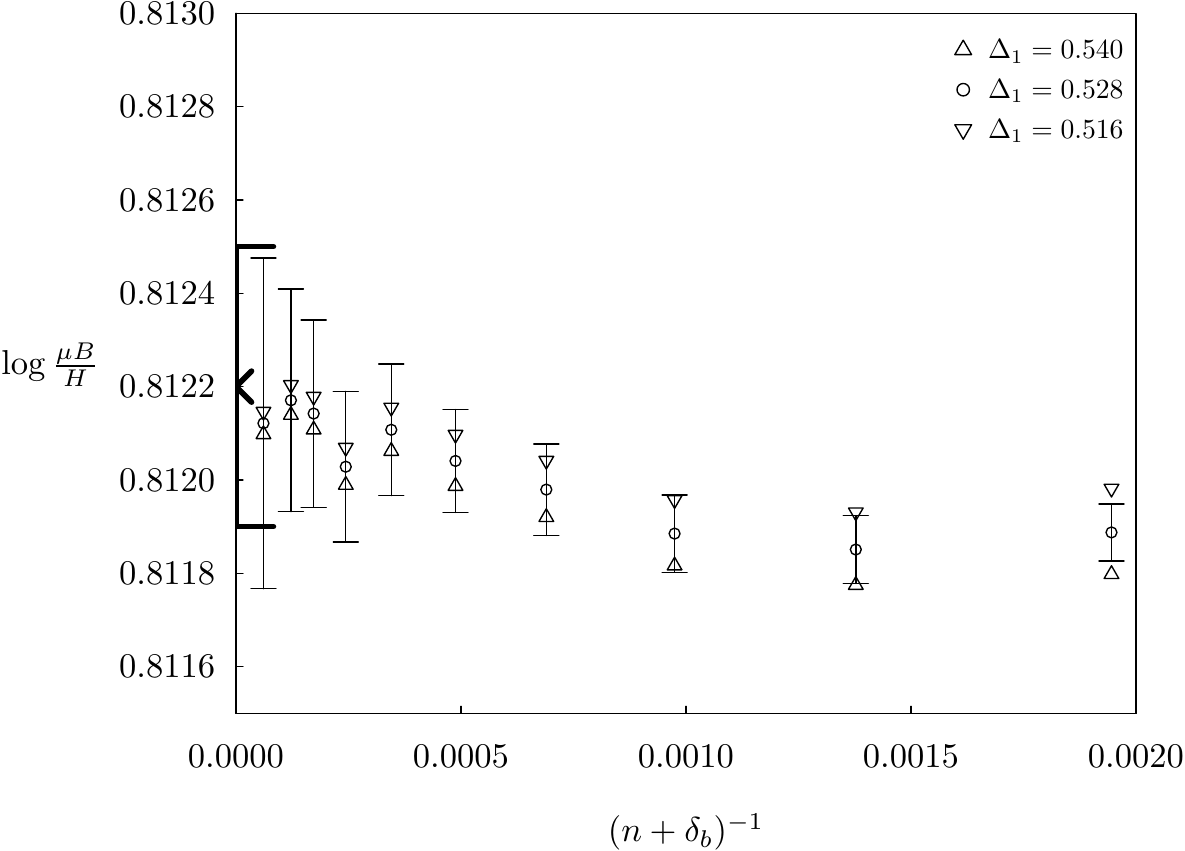}
\end{center}
\vspace{-4ex}
\caption{Estimates of $\mu B/H$ from fits to Monte Carlo
estimates of ratios $b_{n+1}/t_n$, with correction-to-scaling
exponent $\Delta_1 = 0.516, 0.528$, and $0.54$.
Our extrapolated
estimate, $\log \frac{\mu B}{H} = 0.81220(30)$, is shown in the
plot.
Hence
$\frac{\mu B}{H} = 2.25286(68)$.
\label{fig:bh_d3_delta_amplitude}}
\end{minipage}
\end{center}
\end{figure}

The parameters $\delta_h$ and $\delta_b$ were introduced by hand to
reduce the slope of the fits in
Figs.~\ref{fig:hs_d3_delta}-\ref{fig:bh_d3_delta_amplitude}. We found that $\delta_h
= 2$ and $\delta_b = 3$ do this job adequately, which made it easier to
extrapolate our estimates to $n = \infty$.
Effectively, this perturbation provides a crude method of approximately
cancelling out the next-to-leading corrections to scaling.
This trick is not useful for the fits involving the universal amplitude
ratio, as in this case the leading exponent is zero which means the corrections
introduced have small amplitude.

For the critical
exponents corresponding to the amplitude ratio, we performed fits 
using (\ref{eq:log_ratio}) to verify that the combination of critical
exponents is very nearly zero. For the amplitude ratio, we assumed
that the combination of exponents in (\ref{eq:scaling}) is exactly zero, and 
used
(\ref{eq:direct_ratio}) to directly estimate the amplitude.

In each case we have confirmed that the goodness-of-fit is approximately
$1$ in the large $n$ regime,
which indicates that the fitting form is sensible and that sub-leading
terms are lost in the statistical noise.

In Figs~\ref{fig:hs_d3_delta} and \ref{fig:hs_d3_delta_amplitude}
we
plot estimates of exponents and amplitudes from (\ref{eq:log_chi_h}),
obtaining estimates for $\gamma_1 - \gamma$ and $\frac{H}{A}$.
In Figs~\ref{fig:bh_d3_delta} and \ref{fig:bh_d3_delta_amplitude}, we
plot estimates of exponents and amplitudes from (\ref{eq:log_chi_b}),
obtaining estimates for $\gamma_b - \gamma_1$ and $\frac{\mu B}{H}$.
Note that the scaling relation in (\ref{eq:scaling}) implies that  
$\gamma_1 - \gamma = \gamma_b - \gamma_1$.

\begin{figure}[htb]
\begin{center}
\includegraphics[width=0.6\textwidth]{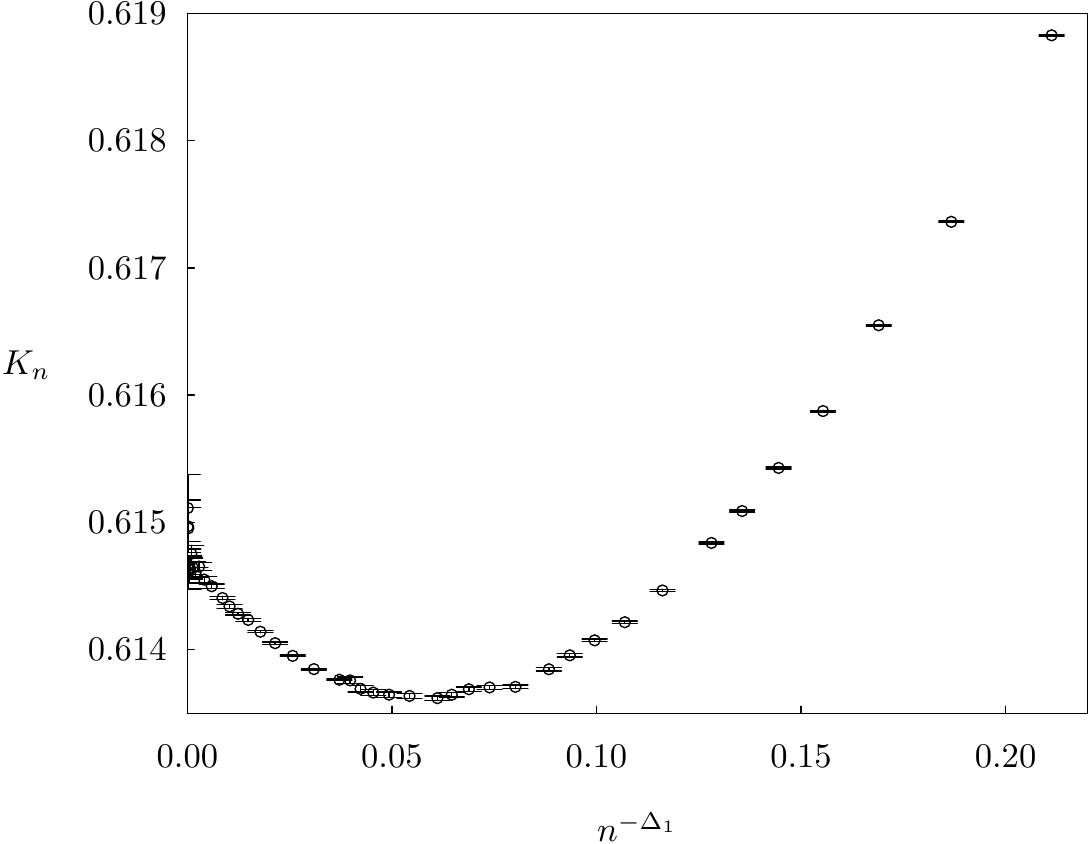}
\end{center}
\vspace{-4ex}
\caption{Plot of $K_n = t_n^2/(b_{n+1} c_n)$ for $n \geq 20$ to show the
strong next-to-leading corrections to scaling. It is impossible to
reliably extrapolate these data for $n \lessapprox 500$ without taking
the next-to-leading corrections into account. 
\label{fig:h2bs_d3_full_amplitude}}
\end{figure}

In Fig.~\ref{fig:h2bs_d3_full_amplitude} we plot the universal amplitude
ratio as a function of $n$, to illustrate the grave difficulty in
extracting reliable estimates from short series. In this case no fits
are performed, we simply plot the amplitudes against the leading
correction to scaling which corresponds to $\Delta_1$. One can see that the
sub-leading corrections to scaling are extremely large, and overwhelm
the leading $\Delta_1$ corrections to scaling until $n$ is of the order
of 500 or so.
To be able to extract reliable estimates from the low order coefficients
one must therefore take into account the next-to-leading corrections. 

\begin{figure}[htb]
\begin{center}
\begin{minipage}{0.45\textwidth}
\begin{center}
\includegraphics[width=1.0\textwidth]{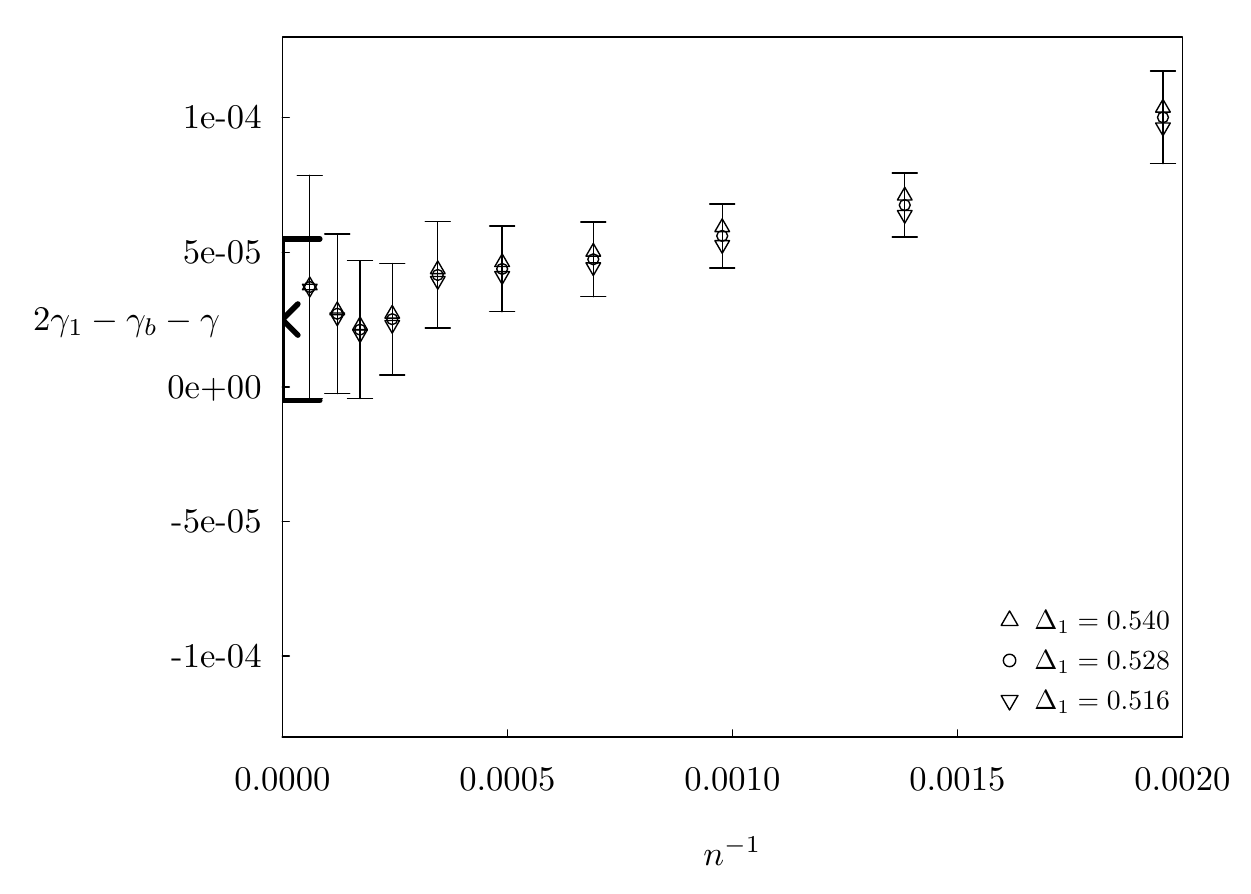}
\end{center}
\vspace{-4ex}
\caption{Estimates of $2\gamma_1 - \gamma_b - \gamma$ from fits to Monte Carlo
estimates of ratios $K_n = t_n^2/(b_{n+1} c_n)$, with correction-to-scaling
exponent $\Delta_1 = 0.516, 0.528$, and $0.54$.
Our extrapolated
estimate is $2\gamma_1 - \gamma_b - \gamma = 0.000025(30)$.
\label{fig:h2bs_d3_delta}}
\end{minipage}
\hspace{2em}
\begin{minipage}{0.45\textwidth}
\begin{center}
\includegraphics[width=1.0\textwidth]{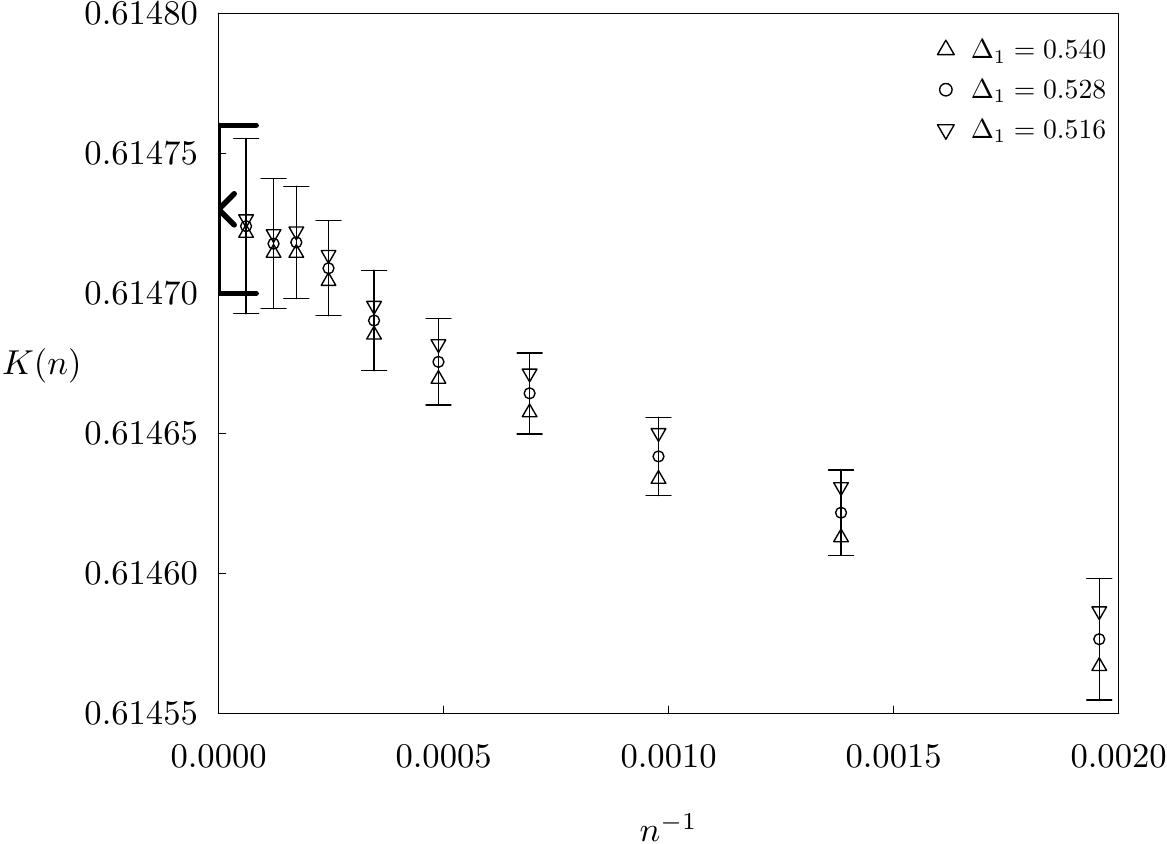}
\end{center}
\vspace{-4ex}
\caption{Plot of amplitude estimates from fits to Monte Carlo estimates
of $K_n = t_n^2/(b_{n+1} c_n)$,
biased to have $2\gamma_1 - \gamma_b - \gamma = 0$, and 
with correction-to-scaling
exponent $\Delta_1 = 0.516, 0.528$, and $0.54$.
Our extrapolated
estimate is $K = 0.614730(30)$.
\label{fig:h2bs_d3_delta_amplitude}}
\end{minipage}
\end{center}
\end{figure}

In Fig.~\ref{fig:h2bs_d3_delta}, we show our 
estimates of $2\gamma_1 - \gamma_b - \gamma$ from fitting
(\ref{eq:log_ratio}). The fits strongly suggest that this combination of
exponents is indeed identically zero.
In Fig.~\ref{fig:h2bs_d3_delta_amplitude}, we show our fits of
(\ref{eq:direct_ratio}) with $2\gamma_1 - \gamma_b -
\gamma$ biased to be zero, i.e. we assume that the scaling relation
(\ref{eq:scaling}) is
correct. By doing this we can obtain far more accurate fits of this
putative universal amplitude ratio than we were able to for the other amplitude
ratios.

\subsection{Summary of results}

We summarise our Monte Carlo results in Table~\ref{tab:estimates}, providing
comparisons between different estimates of $2\gamma_1 - \gamma_b -
\gamma$ and $K$
from fitting data from the two 
computer experiments separately, and together. Reassuringly, we find in
each case that the confidence intervals overlap. Note that the direct
estimate of $K$ is an order of magnitude more accurate than the combined
estimate, due to the biasing of the exponents to satisfy
(\ref{eq:scaling}).

\begin{table}[!ht]
\begin{center}
\bgroup
\def\arraystretch{1.6}
\setlength{\tabcolsep}{3em}
\begin{tabular}{clc}
\hline
Quantity & Estimate & Source \\
\hline
$\gamma_1 - \gamma$ & -0.479290(15) & Fig.~\ref{fig:hs_d3_delta} \\
$\gamma_b - \gamma_1$ & -0.479315(20) & Fig.~\ref{fig:bh_d3_delta} \\
$2\gamma_1 - \gamma_b - \gamma$ & 0.000025(25) & 
Fig.~\ref{fig:hs_d3_delta} and Fig.~\ref{fig:bh_d3_delta} \\
$2\gamma_1 - \gamma_b - \gamma$ & 0.000025(30) & Fig.~\ref{fig:h2bs_d3_delta} \\
$\frac{H}{A}$ & 1.38431(41) & Fig.~\ref{fig:hs_d3_delta_amplitude}\\
$\frac{\mu B}{H}$ & 2.25286(68) & Fig.~\ref{fig:bh_d3_delta_amplitude}\\
$K$ & 0.61447(26) & 
Fig.~\ref{fig:hs_d3_delta_amplitude} and Fig.~\ref{fig:bh_d3_delta_amplitude} \\
$K$ & 0.614730(30) &
Fig.~\ref{fig:h2bs_d3_delta_amplitude} \\
\hline
\end{tabular}
\egroup
\caption{Summary of estimates made via analysis of Monte Carlo data.}
\label{tab:estimates}
\end{center}
\end{table}

In addition, we combine our direct Monte Carlo estimates with previous
exponent estimates $\gamma = 1.156957 \pm
0.000009$~\cite{C14} and $\nu  = 0.587597 \pm 0.000007$~\cite{C10}. 
We then obtain estimates for the TAW
exponent $\gamma_1 = (\gamma_1 - \gamma) + \gamma = -0.479290(15) +
1.156957(9) = 0.677667(17)$, and the bridge exponent
$\gamma_b = (\gamma_b - \gamma_1) + (\gamma_1 - \gamma) + \gamma =
-0.479315(20) - 0.479290(15) + 1.156957(9) = 0.198352(27)$. N.B., we
combine the confidence intervals as if they were independent statistical
estimates of a single standard deviation.

The Monte Carlo estimates are consistent with the series analysis estimates of
Sec.~\ref{sec:seriesextension} ($\gamma_{1} = 0.676 \pm 0.002$ and
$\gamma_b = 0.199 \pm 0.002$) but are dramatically more precise.

From the scaling relation (\ref{eq:scalingb}), we calculate $\gamma_{11}
= \gamma_b - \nu = 0.198352(27) - 0.587597(7) = -0.389245(28)$.

From the scaling relation (\ref{eq:bscaling}), we calculate 
$b = -\gamma_b/(2\nu) + d/2 = 1.331218(23)$, which may be compared with
the direct estimate due to Kennedy of $b = 1.3303(3)$~\cite{K15}.

For the exponent $\theta$, the hand-waving argument in
Sec.~\ref{sec:theta} gave
us the scaling relation (\ref{eq:thetascaling}), from which we obtain
$\theta = 1 - \gamma_b/\nu = 0.662435(46)$.
This indirect estimate of $\theta$ 
should be compared with the direct estimate of $\theta = 0.7 \pm
0.1$ and the indirect estimate of $\theta = 0.661 \pm 0.004$ from the series
analysis in Sec.~\ref{sec:theta}.

We summarise these results in Table~\ref{tab:summary}.
In each case, our estimates improve significantly on those previously
available in the literature.

\begin{table}[!ht]
\begin{center}
\bgroup
\def\arraystretch{1.6}
\setlength{\tabcolsep}{3em}
\begin{tabular}{ll}
\hline
Quantity &    Estimate  \\
\hline
$\gamma_1$ &  0.677667(17)      \\
$\gamma_b$ &  0.198352(27)           \\
$\gamma_{11}$ & -0.389245(28)            \\
$b$ &   1.331218(23)     \\
$\theta$ &  0.662435(46)  \\
\hline
\end{tabular}
\egroup
\caption{Summary of estimates for critical exponents from Monte Carlo.}
\label{tab:summary}
\end{center}
\end{table}

\section{Discussion and conclusion}
\label{sec:discussion}

\subsection{Possible improvements and extensions}

For the enumerations, we utilised a straight-forward backtracking
algorithm despite the fact that some improved algorithms have been
developed in recent years for self-avoiding walks for $d=3$. 
The three innovations are the use of the lace expansion and the two-step
method~\cite{CLS07}, and the length-doubling
algorithm~\cite{SBB11}.

Of these methods, it is unlikely that the lace expansion would help as
this works by enumerating configurations like self-avoiding polygons
instead of SAWs, and there would not be as much of a gain for 
TAWs and bridges.

The length-doubling algorithm is extremely promising, but would be
challenging to adapt to the present problem. If this could be done, then
it may be possible to extend the series past the existing $n=36$ for
SAWs~\cite{SBB11}.

Finally, the two-step method could be adapted quite easily and 
used to increase the length of the series
for the same computational effort, perhaps by a few terms.

Unfortunately it is unlikely that the
extent of the improvement would be sufficient to be competitive with 
Monte Carlo, even in the case of the length-doubling algorithm.

To improve the Monte Carlo results it would be possible to perform the
same computer experiment with weakly self-avoiding walks (also known as
the
Domb-Joyce model) where the weight factor is chosen so that the leading
correction-to-scaling term corresponding to $\Delta_1$ has small amplitude.

In future we will perform further Monte Carlo computer experiments 
for a variety of two- and three-dimensional lattices to
test our hypothesis that $K$ is indeed a universal amplitude ratio.
We have some preliminary numerical evidence that this ratio is the same
for the simple cubic, face-centred cubic, and body-centred cubic lattices.

\subsection{Conclusion}

We have obtained high precision estimates of a variety of exponents and
amplitudes associated with terminally attached self-avoiding walks and
bridges in three dimensions. In particular, we have confirmed to high
precision that the scaling relation in (\ref{eq:scaling}) holds, and
estimated the corresponding amplitude ratio $K$ which we believe to be
universal.

\FloatBarrier

\appendix


\section{Enumerations}
\label{sec:enumerations}

\begin{table}[htbp]
\begin{center}
\begin{tabular}{|p{30pt} p{110pt} p{110pt} |}
\hline
$n$ &TAWs & Bridges \\
\hline
0 &            1                       & 0                      \\
1 &            5                       & 1                      \\
2 &            21                      & 5                      \\
3 &            93                      & 21                     \\
4 &            409                     & 89                     \\
5 &            1853                    & 369                    \\
6 &            8333                    & 1553                   \\
7 &            37965                   & 6573                   \\
8 &            172265                  & 28197                  \\
9 &            787557                  & 122093                 \\
10 &           3593465                 & 533369                 \\
11 &           1647784                 & 2345429                \\
12 &           7548110                 & 10366677               \\
13 &           346960613               & 46013585               \\
14 &           1593924045              & 204927833              \\
15 &           7341070889              & 915448621              \\  
16 &           33798930541             & 4100092693             \\
17 &           155915787353            & 18407472565            \\
18 &           719101961769            & 82815889677            \\
19 &           3321659652529           & 373321398437           \\
20 &           15341586477457          & 1685838489629          \\
21 &           70944927549085          & 7625255897889          \\
22 &           328054694768261         & 34541044018277         \\
23 &           1518490945278377        & 156678876463321        \\
24 &           7028570356547189        & 711593257794069        \\
25 &           32560476643826933       & 3235634079777801       \\
26 &           150838831585499069      & 14728414578753489      \\
27 &                                   & 67110197685388181      \\
28 &                                   & 306074586987649389     \\
\hline
\end{tabular}
\caption{Simple cubic lattice TAWs and bridges of length
$n$.}
\label{tab:one}
\end{center}
\end{table}

\begin{sidewaystable}
\begin{center}
\scalebox{0.85}{
\begin{tabular}{|cccccccccc|}
\hline
$n$ &height 1 &height 2&height 3&height 4&height 5&height 6&height
7&height 8 & height 9 \\
\hline
1& 1 &&&&&&&&\\
2& 4 &&&&&&&&\\
3& 12 &&&&&&&&\\
4& 36 &&&&&&&&\\
5& 100 & &&&&&&&\\
6& 284 & 12 &&&&&&&\\
7& 780  & 144 &&&&&&&\\
8& 2172 &1104 &&&&&&&\\
9& 5916&6744 & 12 &&&&&&\\
10& 16268 &35484& 252  &&&&&&\\
11& 44100 &170344& 3332 &&&&&&\\
12& 120292&760656& 33188& 12  &&&&&\\
13& 324932 &3240024& 272056& 360 &&&&&\\
14& 881500 & 13266260 & 1944216& 6708&&&&&\\
15& 2374444& 52897744& 12469228& 92040& 12  &&&&\\
16& 6416596& 206110864 & 73725500& 1023008& 468&&&&\\
17& 17245332 & 790696704& 407848052& 9733040& 11232&&&&\\
18& 464666767& 2990802868&  2141811776&  81770392 & 195756& 12    &&&\\
16& 124658732& 11206740432&  10770243016& 622677008& 2739436& 576&&&\\
20& 335116620 & 41617757096&  52310275276& 4368861352& 32483296&
16904&&&\\
21& 897697164  & 153655478264& 246718702580& 28658873096& 337104536&
356792& 12 & &\\
22& 2408806028 & 563984887964& 1136186599756& 177561158756& 3140961204&
6011640& 684&&\\
23& 6444560484 & 2062508494408& 5126606123704& 1048578720568&
26724562112& 85255000& 23724&&\\
24& 17266613812 &  7512897046320& 22746305294180& 5942315016020&
210656680128& 1052099212 &587604& 12&\\
25& 46146397316& 27302951014488& 99466683163980& 32513181993504&
1554742879184& 11586414632& 11569024& 792& \\
26& 123481354908 & 98952488047860& 429730858249952&
172562782519172&10843077619792&  115848555164&  190791832& 31692& \\
27&329712786220 &358096887158816& 1837073895696056& 892191489850824 &
71975965355296 & 1066811813096 & 2726830756& 900648 & 12 \\
28& 881317491628 & 1293405464647968 & 7784187577745060 &4508562016424432
&457618567787368 & 9145657703324 & 34630365164 & 20280096 & 900 \\

\hline
\end{tabular}}
\caption{Simple cubic lattice irreducible bridges of length $n$ and
various heights.}
\label{tab:two}

\end{center}
\end{sidewaystable}

\FloatBarrier

\section{Monte Carlo data}
\label{sec:mcdata}

\begin{table}[!htb]
\begin{center}
\begin{tabular}{rll}
\hline
$n$      &  \multicolumn{1}{c}{$\langle \chi_h \rangle_n$} & \multicolumn{1}{c}{$\langle \chi_b
\rangle_{\mathcal{H}_n}$} \\
\hline
511      & 0.06923625(26)   & 0.1128060(11)  \\
723      & 0.05870409(26)   & 0.0956331(11)  \\
1023     & 0.04975890(24)   & 0.08104698(73) \\
1447     & 0.04217388(24)   & 0.0686814(10)  \\
2047     & 0.03573665(22)   & 0.05818965(67) \\
2895     & 0.03028231(23)   & 0.04930093(93) \\
4095     & 0.02565517(21)   & 0.04176450(63) \\
5791     & 0.02173570(21)   & 0.03538056(85) \\
8191     & 0.01841284(19)   & 0.02996852(60) \\
16383    & 0.01321296(18)   & 0.02150202(57) \\
32767    & 0.00948020(16)   & 0.01542623(53) \\
65535    & 0.00680166(15)   & 0.01106585(50) \\
131071   & 0.00487920(13)   & 0.00793876(47) \\
262143   & 0.00350038(12)   & 0.00569498(44) \\
524287   & 0.00251118(12)   & 0.00408485(40) \\
1048575  & 0.001801345(94)  & 0.00293075(36) \\
2097151  & 0.001292118(84)  & 0.00210227(34) \\
4194303  & 0.000926791(74)  & 0.00150794(30) \\
8388607  & 0.000665041(66)  & 0.00108145(26) \\
16777215 & 0.000477001(60)  & 0.00077565(24) \\
33554431 & 0.000342261(50)  & 0.00055642(23) \\
\hline
\end{tabular}
\end{center}
\caption{Estimates of $\langle \chi_h \rangle_n$ and  $\langle \chi_b
\rangle_{\mathcal{H}_n}$ from Monte Carlo computer experiments.\label{tab:mcbridge}}
\end{table}

\FloatBarrier

\section*{Acknowledgements}
NC wishes to thank the Australian Research Council for supporting this
work under the Future Fellowship scheme (project number FT130100972).
AJG wishes to thank the Australian Research Council for supporting this
work through grant DP120100931.

\end{document}